\documentclass{ws-ijmpd}
\usepackage{aas_macros}
\usepackage{graphicx}
\usepackage{hyperref}
\begin{document}

\markboth{Remo Ruffini}
{Selected topics}

%
\catchline{}{}{}{}{}
%

\title{Selected topics on: \\
1) proposal of interpreting the Crab supernova with a GRB \\
2) progress in identifying the seven GRBs episodes \\
3) the role of Sagittarius A in identifying the dark matter component (the X fermion)}

\author{R.~Ruffini, C.~Sigismondi, Y.~Wang, H.~Quevedo, S.~Zhang, Y.~Aimuratov, P.~Chardonnet, C.L.~Fryer, T.~Mirtorabi, R.~Moradi, M.~Prakapenia, F.~Rastegarnia, S.-S.~Xue}

\address{ICRANet, Piazza della Repubblica 10, I-65122 Pescara, Italy\\ICRANet, 1 Avenue Ratti, 06000 Nice, France\\ICRA, Dipartimento di Fisica, Sapienza Universit\`a di Roma, Piazzale Aldo Moro 5, I-00185 Roma, Italy\\INAF, Viale del Parco Mellini 84, 00136 Rome, Italy\\
$^*$E-mail: ruffini@icra.it\\}

\maketitle

\begin{history}
\received{Day Month Year}
\revised{Day Month Year}
\end{history}

\begin{abstract}
As the fiftieth anniversary of our common effort in the field of relativistic astrophysics is approaching, we offer a new look to some of our acquired knowledge in a more complete view, which evidence previous unnoticed connections. They are gaining due prominence in reaching a more complete picture evidencing the main results.

We outline the history of GRB observations along with a summary of the contributions made by our group to develop the BdHN interpreting model. We show the seven Episodes characterizing the most powerful BdHNe I occurred to date: GRB 190114C and GRB 220101A. New inferences for the explanation of the highest energy radiation in the TeV
are presented.
\end{abstract}


\ccode{PACS numbers:}

\section*{\Large Proposal of Interpreting the Crab supernova with a GRB}

\section{Introduction}\label{sec:introduction}
\begin{figure}
    \centering
    \includegraphics[width=0.5\linewidth]{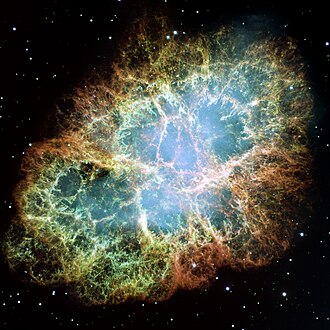}
    \caption{The filamentary structure of the synchrotron emission from the Crab Nebula. 
    Credit: NASA/ESA.}
    \label{fig:crab_nebula}
\end{figure}

There is no better way to start a presentation on Relativistic Astrophysics with a splendid picture of the Crab Nebula remnant of the Supernova (SN) which occurred on 4th of July 1054. Here is the image taken in 2005 by the Hubble Space Telescope (see Fig.~\ref{fig:crab_nebula}) still expanding today at speed $v\approx200$~km/s.

{}

Thousands of SNe have been discovered, since the early historical ones and a vast literature exists with expert scientists in the field creating expert groups dedicated to record the SN properties, see, e.g., Supernova Remnant:

\begin{itemize}
    \item \textit{``Supernovae''} by Iosif Shklovskii (1968);\cite{{1969supe.book.....S}}
    \item \textit{``Supernovae and Nucleosynthesis''} by David Arnett (1996); \cite{Arnett1996}
    \item \textit{``Physics and Evolution of Supernova Remnants''}  by Jacco Vink (2020). \cite{Vink2020}
\end{itemize}

\begin{figure}
    \centering
    \includegraphics[width=1\linewidth]{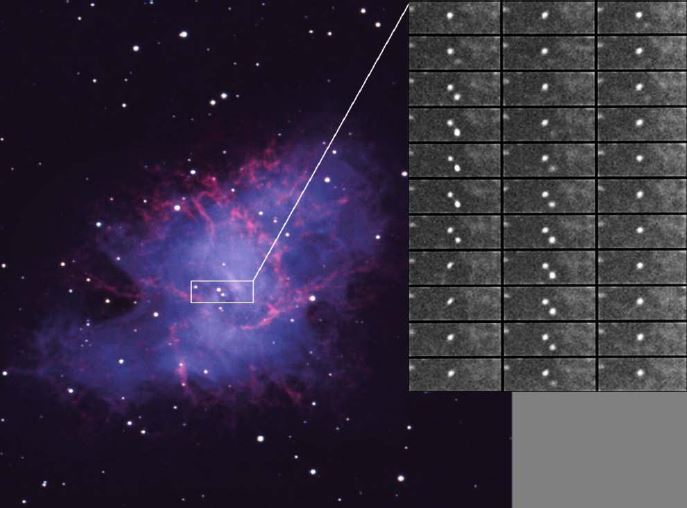}
    \caption{The time varying emission of the star NP0532 turning on and off with characteristic period of $33$~ms. Credit: N.A.Sharp/AURA/NOAO/NSF.}
    \label{fig:ms_pulsar}
\end{figure}

The next fundamental observation has been associated with the discovery inside the Crab Pulsar of a $33$~ms pulsar as first reported by Hewish et al. (1968) \cite{Hewish1968} (see Fig.\ref{fig:ms_pulsar})

Also in this case since that discovery, thousands of pulsars have been discovered and a large group of experts and publications have developed a well identified independent field of scientific research:

\begin{itemize}
    \item \textit{``Handbook of Pulsar Astronomy''} by Lorimer and Kramer (2005) \cite{Lorimer-Kramer2004};
    \item \textit{``Pulsar Astronomy''} by Lyne and Graham-Smith (2012) \cite{Lyne-Graham2012};
    \item Recently discovered more than 1000 pulsars by FAST\\ (\url{http://zmtt.bao.ac.cn/GPPS/}).
    \end{itemize}
    
\begin{figure}
    \centering
    \includegraphics[width=1\linewidth]{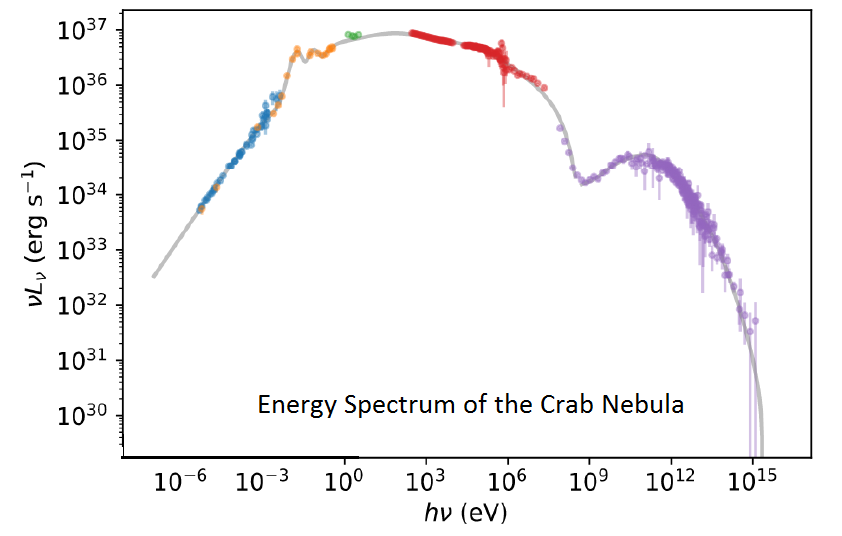}
    \caption {The spectra of the Crab Nebula obtained by vast number of instruments, given in Table~\ref{tab:crab_obs_instruments}.}
    \label{fig:crab_spectra}
\end{figure}

The spectral observations of the Crab SN Remnant represented in Fig.~\ref{fig:crab_nebula}, have gained the attention of an extremely vast community of astronomers observing in the radio, in the optical, in the X-ray and High energy extending to the GeV and to the PeV. A totally independent group of scientists have developed the technological most advanced observational techniques from space, from the ground and underground also involved a very large community of scientists using more than thousand instruments of multi-wavelength range (see Tab.~\ref{tab:crab_obs_instruments}), and finally reaching a global description of the spectral properties in more than $5000$~publications. The observations range from the optical, to the X-ray, to the GeV to the PeV and a unified theoretical explanation is still to be reached (see Fig.~\ref{fig:crab_spectra}).

\begin{table}
    \centering
    \begin{tabular}{|l|l|} 
    \hline AGILE - Astro-rivelatore Gamma a Immagini LEggero          \\ 
    \hline ALMA - Atacama Large Millimeter/submillimeter Array        \\ 
    \hline BeppoSAX - Satellite per Astronomia X                      \\
    \hline CFHT - Canada-France-Hawaii Telescope                      \\ 
    \hline Chandra - Chandra X-ray Observatory                        \\ 
    \hline Fermi - Fermi Gamma-ray Space Telescope                    \\ 
    \hline GBT - Green Bank Telescope                                 \\ 
    \hline H.E.S.S. - High Energy Stereoscopic System                 \\ 
    \hline HSO - Herschel Space Observatory                           \\ 
    \hline HST - Hubble Space Telescope                               \\ 
    \hline Integral - INTErnational Gamma-Ray Astrophysics Laboratory \\ 
    \hline JCMT - James Clerk Maxwell Telescope                       \\
    \hline JVLA - Karl G. Jansky Very Large Array                     \\ 
    \hline Keck - Keck Observatory                                    \\
    \hline LOFAR - Low-Frequency Array                                \\
    \hline MAGIC - Major Atmospheric Gamma Imaging Cherenkov Telescopes \\
    \hline NuSTAR - Nuclear Spectroscopic Telescope Array \\
    \hline ROSAT - Röntgensatellit \\
    \hline SOFIA - Stratospheric Observatory for Infrared Astronomy\\
    \hline Spitzer - Spitzer Space Telescope\\
    \hline Subaru - Subaru Telescope\\
    \hline Swift - Swift Gamma-Ray Burst Mission\\
    \hline VLA - Very Large Array\\
    \hline VLT - Very Large Telescope\\
    \hline VERITAS - Very Energetic Radiation Imaging Telescope Array System\\
    \hline XMM-Newton - X-ray Multi-Mirror Mission-Newton\\
    \hline
    \end{tabular}
    \caption{Tab. 1. The multi-messenger instruments which have participated in determining the overall spectra and luminosity of the Crab Nebula, expressed in Figure~\ref{fig:crab_spectra}.}
    \label{tab:crab_obs_instruments}
\end{table}

\begin{figure}
    \centering
    \includegraphics[width=0.5\linewidth]{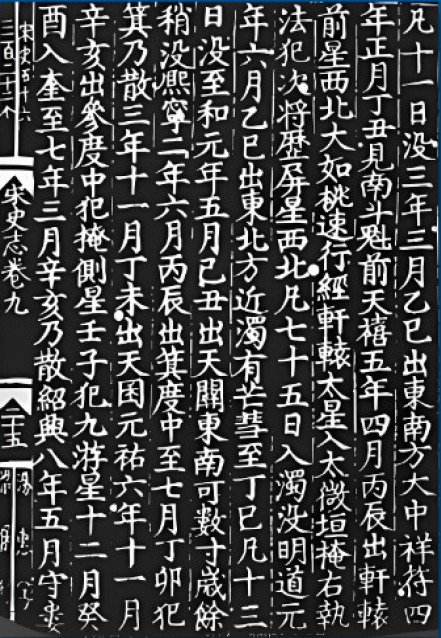}
    \caption {The oldest and most detailed accounts of the SN 1054 are from Song Huiyao and Song Shi, with the apparition's date of 4 July. This figure was reproduced from the book ``Gravitation'' by C.W.~Misner, K.S~Thorne, J.A.~Wheeler.}
    \label{fig:sn_1054_document}
\end{figure}

At the same time, the astronomers working on the history of the initial SN explosion and identification of associated historical events, have largely expanded their knowledge, see Fig. \ref{fig:sn_1054_document} and Tab.~\ref{tab:sn_1054_notes}.

\begin{figure}
    \centering
    \includegraphics[width=1\linewidth]{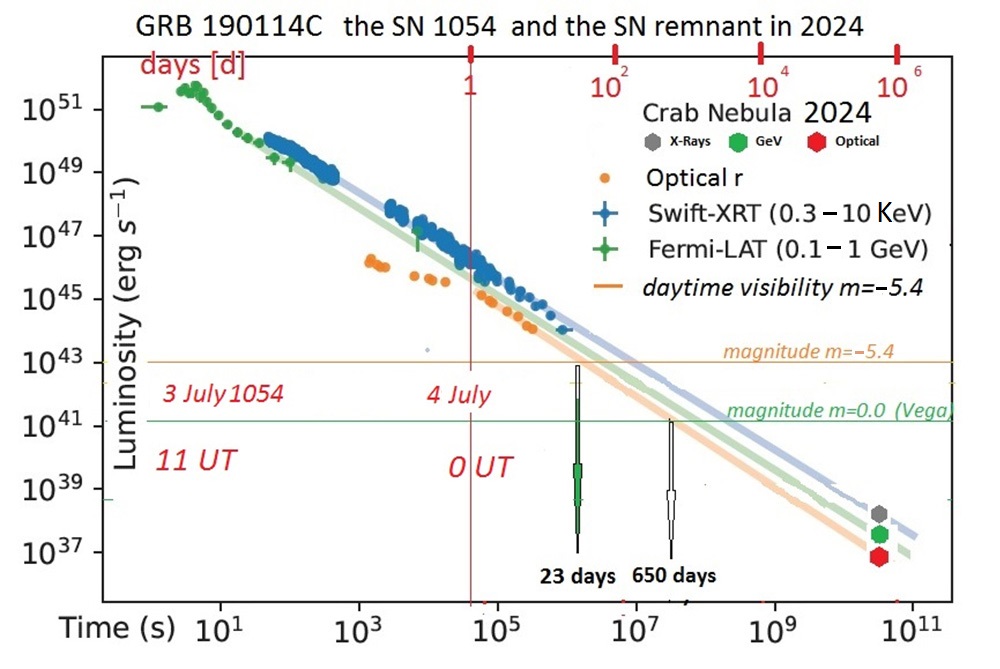}
    \caption{The extrapolation after $971$~year of the GRB~1901114C has been obtained with a common power-law decay  for the optical, X-ray and  GeV afterglows, having assumed a boresight angle of the source within 68 degrees for Fermi LAT data, and further assumed that the beaming angle of the source in the GeV is along the line of sight. See however Fig.~\ref{fig:crab_190114c_imprint}. The comparison with the Crab optical, KeV and GeV current spectrum is made in the Fig.~\ref{fig:crab_190114c_imprint}. The dates of 3rd and 4th of July~1054 are referred with the apparition of the GRB near the rising horizon in America, at the meridian in Constantinoples and Egypt and on the following day in China, see also Fig.~\ref{fig:airmass_at_sn}. This figure is reproduced from Ruffini and Sigismondi (2024).} 
    \label{fig:crab_extrapolation}
\end{figure}

\begin{figure}
    \centering
    \includegraphics[width=\linewidth]{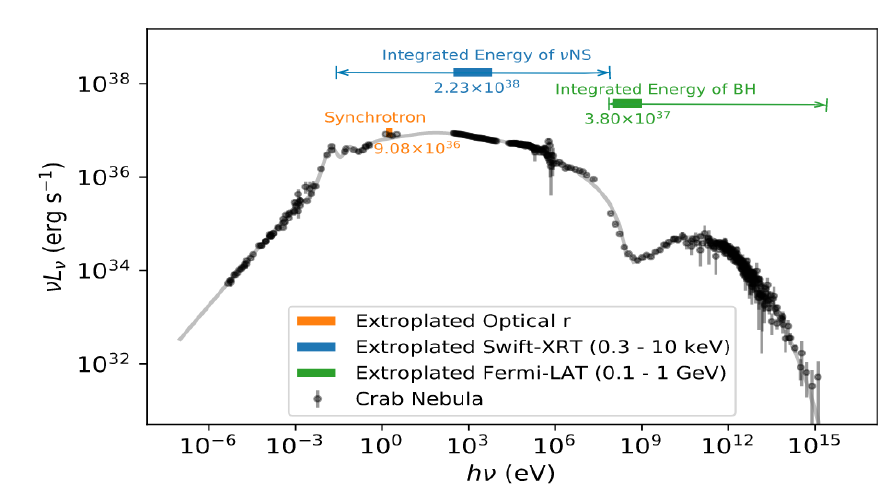}
    \caption{The multi-wavelength spectrum of the Crab Nebula, with afterglow imprint by GRB~190114C.}

    
    \label{fig:crab_190114c_imprint}
\end{figure}

\begin{table}
    \caption{Tab. 2. Historical notes documented the appearance of the SN~1054.}
    \centering
    \begin{tabular}{|c|} 
     \hline {The 1054 Appearance of Supernova (SN)}\\
     \hline {Appeared the 4th of July 1054} \\
     {From the Sung-shih [Annals of the Sung Dynasty] (Astronomical Treatise, chapter 56)}\\
     \hline {Visible in daytime for 23 days} \\
     {Sung-hui-yao [essentials of the Sung dynasty history] (Chapter 52)}\\
     \hline {Visible for 650 days} \\
     {from the Sung-shih [Annals, if the Sung dynasty] (Astronomical Treatise, chapter 9)}\\
     \hline {After the solar conjunction of 27 May, and 6 to 16 days before the 4th July} \\
     {from Japanese Reference, see Iosif Shklovskii \cite{1969supe.book.....S}.} \\ 
     \hline  
    \end{tabular}
    \label{tab:sn_1054_notes}
\end{table}

While all these fields kept developing in the inauguration of MG~XVII a new idea was proposed: in order to understand the real physics underlying a Supernova, all these different aspects of the Crab Supernova should not be addressed separately by different specialized groups which are unable to have the overall understanding of the process.

It is indeed amazing that still today there is not a basic understanding of the SN process based on neutrinos interaction.
An alternative proposal was advanced that the SN 1054 be explained as originating from a GRB similar to GRB~190114C in these introductory remarks.

The above aspects can be explained in term of the ``seven episodes'' characterizing a Binary Driven Hypernova of Type~I: originating from a binary system composed of a CO-core of $10M_{\odot}$ and a NS companion of $2M_{\odot}$. Using the new physics of the ``seven episodes'' to reach the real understanding of the above data and also of the real origin of the Supernova. With the new physics of the seven BdHN I episodes, there is a possibility to reach a working model of a Supernova (see below).

Is it possible from the analysis of GRB~190114C to explain the observations of the SN of 1054? That this is indeed possible, it was shown in Fig~\ref{fig:crab_extrapolation}, and presented in the paper by Ruffini and Sigismondi~(2024).\cite{RR-Sigismondi2024} A great tangible success is that the double peak structure observed in Fig.~\ref{fig:crab_spectra} is indeed explained in the optical, in the X-Ray and in the high energy GeV and PeV by the asymptotic emission of the $v$NS pulsar radiation (in the optical), by the synchrotron radiation of the ``X-ray afterglow'' in the X-ray and by the GeV and PeV radiation emitted by the asymptotic emission of the Black Hole (see Fig.~\ref{fig:crab_190114c_imprint}).

The extrapolated values derived from the optical, GeV and X-rays afterglows of GRB~190114C after 971 years, are in good agreement with the optical data, and in general agreement with the X-ray data which can be still proved with spectral analysis of the synchrotron data. It is clear that the much lower luminosity of the observed GeV - TeV emission with respect to the extrapolated one to present time (Fig. ~\ref{fig:crab_extrapolation}) is due to a different orientation of the BH axis with respect to the line of sight. The position of the BH in the remnant, and the BH axis orientation is clearly very strongly time varying in time in view of its intrinsic angular momentum as well as the  the interaction of the BH within the magnetized filamentary structure  of the nebula, well manifested e.g. by the GeV flare-like  emission.\cite{Tavani2011} Since the localization of the BH by the GeV radiation is clearly out of reach by the present dedicated satellites, it make sense to program a new X-rays space mission on large scale following the Chandra and Einstein Probe missions to follow the X-ray emission of matter accretion of the BH and acquire information of its motion within the remnant. Ongoing discussion on such themes have been started at the MG17 meeting with Wei Jianyan, PI of the Einstein Probe, and they are the basis of the Italian-Chinese collaboration proposals presented by ICRANet.


\begin{figure}
    \centering
    \includegraphics[width=1\linewidth]{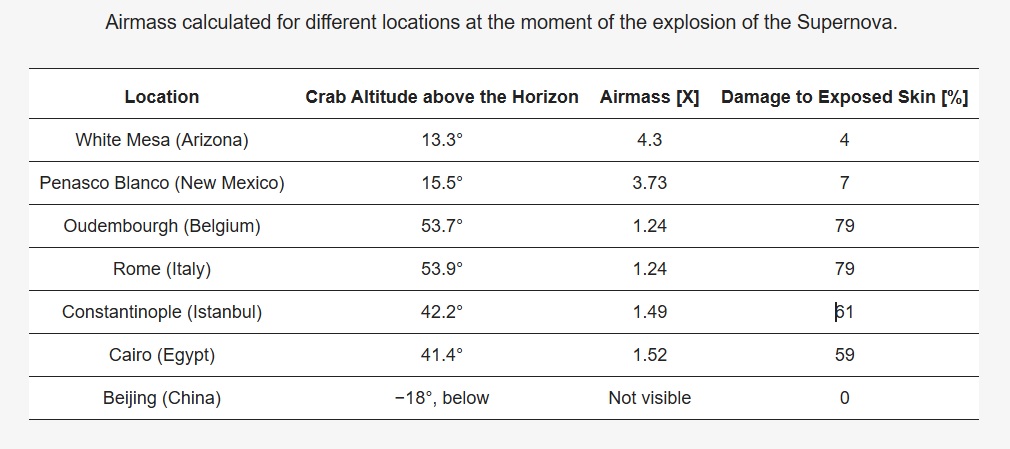}
    \caption {In ancient chronicles, the plagues were connected with the new star's apparition. They correspond with the GRB at low airmasses and in the first half hour. They may have been radiation's plagues. Reproduced from Ruffini and Sigismondi (2024).}
    \label{fig:airmass_at_sn}
\end{figure}

The optical data, coming from the extrapolation of the $v$NS agree remarkably well with the recent observations of the Crab Pulsar, the X-ray data coming from the extrapolation of the X-ray afterglow as well as the high energy extrapolation of the GeV data (see values at $10^{10}$ s in Fig.~\ref{fig:crab_extrapolation}), see also extrapolation in Fig.~\ref{fig:crab_190114c_imprint}, are larger than the observed ones in the Crab \cite{RR-Sigismondi2024,RR-Mirtorabi2025}, this seems to be consistent with a beaming effect of the GeV source in the BdHN I model.\cite{RR-Sigismondi2025}
But in addition to all of this there is also a new conceptual point which needs a complete revolution: in the BdHN~I the SN occurs at the trigger time of the collapse of the CO core and this needs a new understanding of the SN process to be explained within the properties of the CO core. In addition, the existence of the $v$NS has to be identified and duly explained. Similarly an alternative origin to the SN energy is being found on the electrodynamics of a fast rotating core duly endowed with magnetic fields of $10^{6}$~Gauss reaching critical fields and creating the electro positron plasma needed to explain the thermal emission of KeV following in SN rise.\cite{RR-Mirtorabi2025}

\section{Gravitational versus electrodynamical process in the rotational energy extraction from rotating Black Holes following Roy Kerr solutions}\label{sec:gw_em_process}
\subsection{Historical background}\label{sec:historic_backgr}
\begin{figure}
    \centering
    \includegraphics[width=0.5\linewidth]{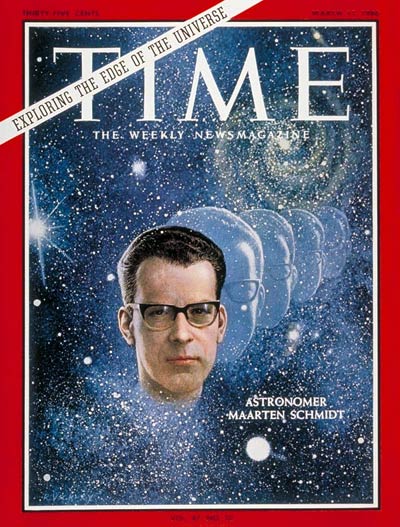}
    \caption{The cover of the Time magazine dedicated to Maarten Schmidt celebrating the discovery of Quasars.}
    \label{fig:time_magazine_cover}
\end{figure}

The most fundamental contribution to General Relativity and Relativistic Astrophysics certainly came from the introduction by Roy Kerr of a solution for a rotating BH, see Kerr~R.P. (PRL, 11, 237, 1963).\cite{Kerr1963} 
This solution was presented at the first Texas Symposium on Relativistic Astrophysics in 1963, following the discovery of Quasar by Maarten Schmidt, see Fig.~\ref{fig:time_magazine_cover} and Schmidt~(Nature, 197, 1040, 1963).\cite{Schmidt1963}

It is interesting that, although many decades have passed, some fundamental building blocks for reaching a satisfactory explanation of quasars are still missing, although some preliminary observational results by James Webb Space Telescope and the work of Luis Ho (see Zhang~Z.~(2025)\cite{Zhang2025} and also invited lecture by Prof. Ho on the 7th Galileo Xu-Guangqui meeting (in press), which held in ICRANet Hq in Pescara from July 7 to 11, 2025) are attracting attention worldwide and let us hope for a positive outcome in the near future.\cite{Wang2025inpress} 
The next fundamental discovery, following the Quasars discovery, came from the Pulsars Discovery (see Fig.~\ref{fig:ms_pulsar}) and from the discovery of GRBs, which are unveiling seven different Episodes of the fundamental physics, which are likely allowing to reach the first working model of a supernova (see also \cite{2509.06243v1}).

\begin{figure}
    \centering
    \includegraphics[width=1\linewidth]{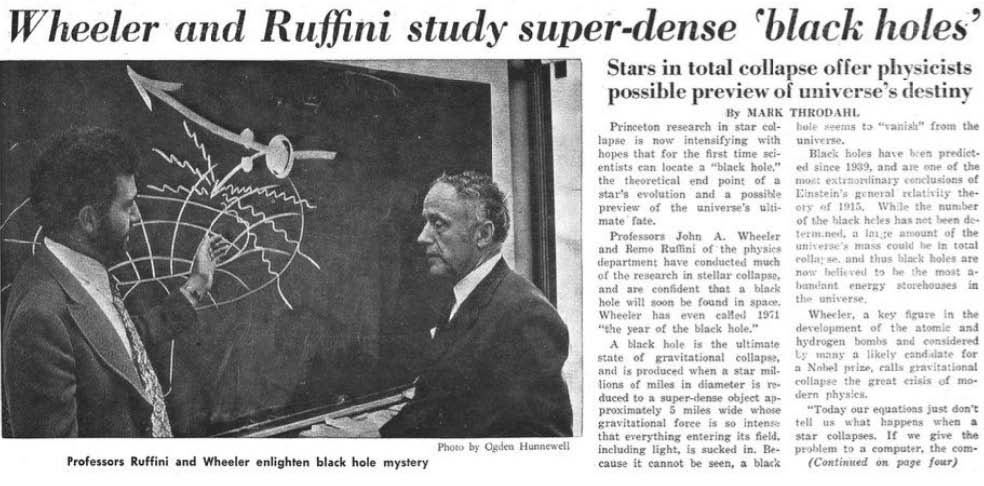}
    \caption{The first page of ``The Princetonian'', the independent daily student newspaper of the Princeton University.}
    \label{fig:princetonian}
\end{figure}

\begin{figure}
    \centering
    \includegraphics[width=0.5
    \linewidth]{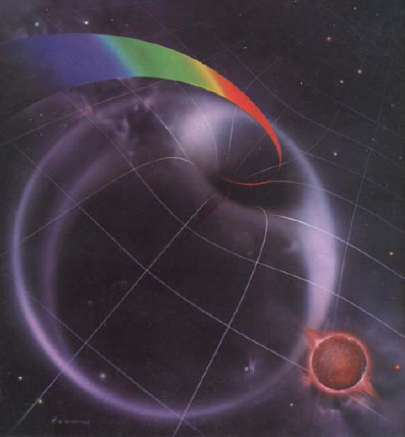}
    \caption{The painting of the BH made H.~Wimmer which was used on the cover of Physics Today, January 1971. Reproduced from Ruffini and Wheeler (Physics Today, 24, 30, 1971)}.
    \label{fig:bh_painting}
\end{figure}

The Kerr solution gave an opportunity to Remo Ruffini and John A. Wheeler (see Fig.~\ref{fig:princetonian}) to formulate the basic concept of a BH, manifested in Physics Today (Vol. 24, 30, 1971),\cite{RR-Wheeler1971} see enclosed the cover image in Fig.~\ref{fig:bh_painting}. This work has reached the current worldwide interest.

\newpage
\subsection{Is there a role of BH for the developing extra-terrestrial civilization?}\label{sec:semi-fiction_approach}
For a while various personalities have addressed such questions, notoriously, Freeman Dyson in his concept of Dyson spheres (1960).\cite{Dyson1960}



\begin{figure}
    \centering
    \includegraphics[width=1.1\linewidth]{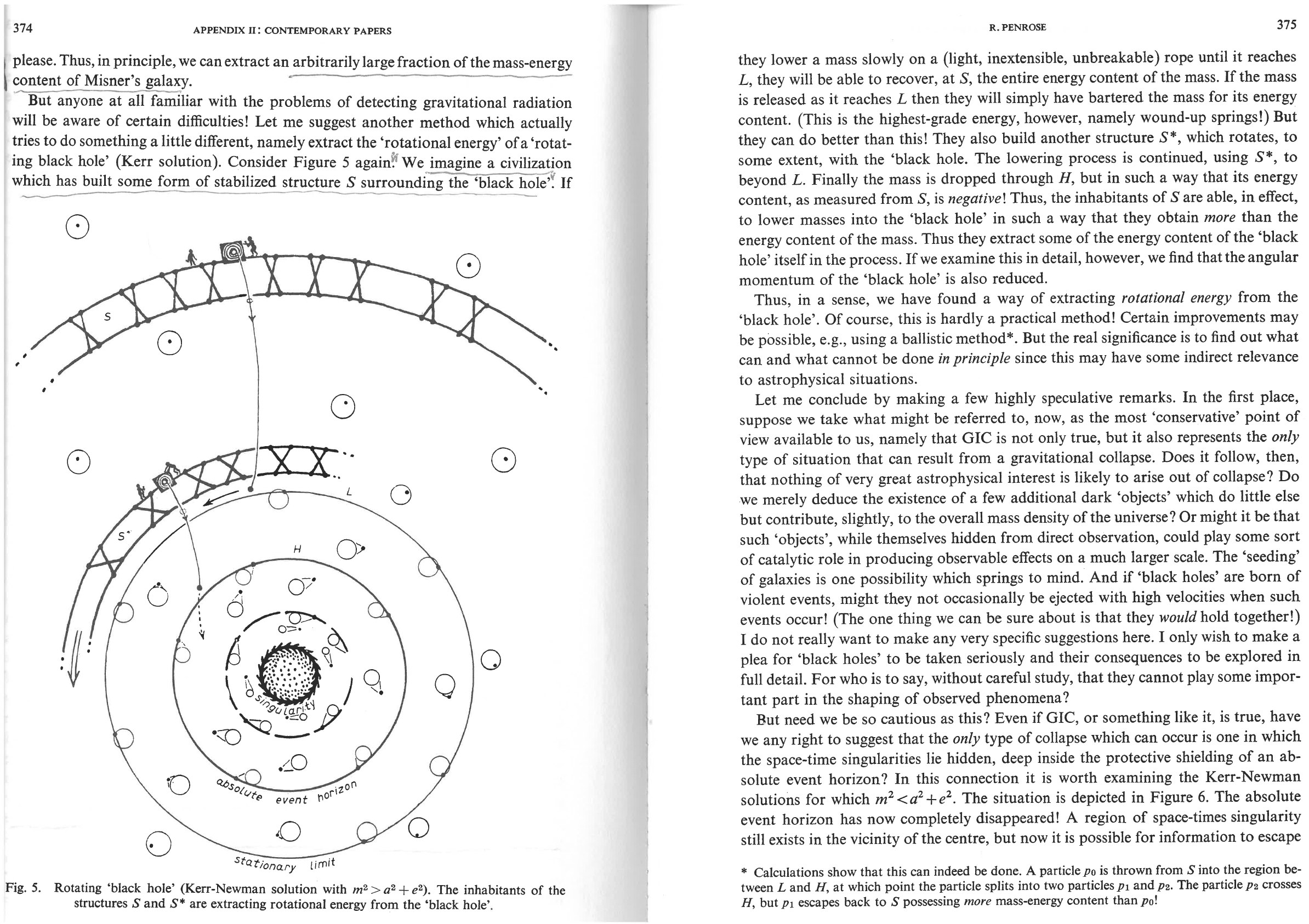}
    \caption{Reproduced from Penrose (1969).}
    \label{fig:penrose_article}
\end{figure}

In parallel to those works, a series of articles called general attention on the possibility of inhabitants of an advanced civilization extracting energy from a Kerr-Newman Black Hole, see Fig.~\ref{fig:penrose_article} from Penrose (1969).\cite{Penrose1969}

\begin{figure}
    \centering
    \includegraphics[width=0.8\linewidth]{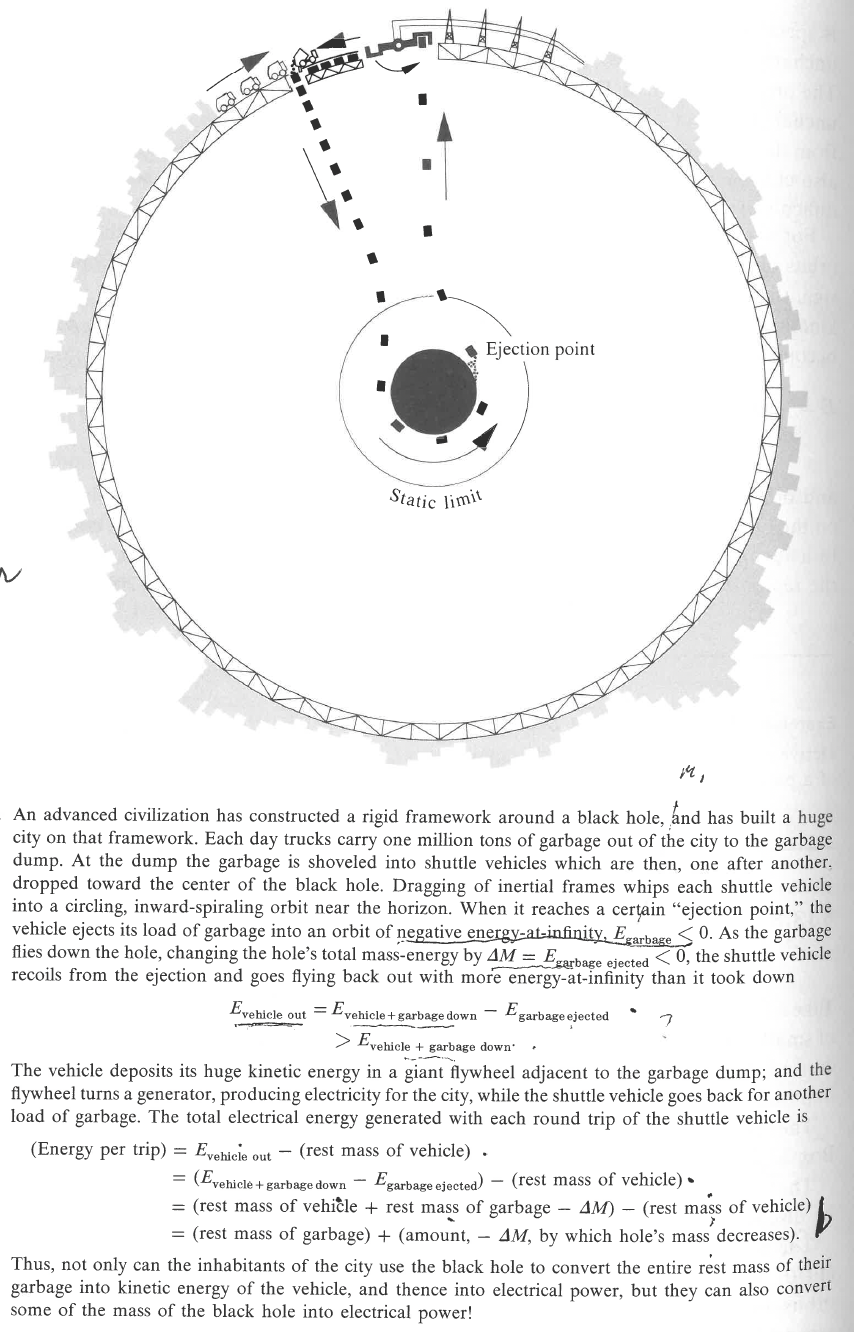}
    \caption{The textbook problem. Reproduced from the book ``Gravitation'' by C.W.~Misner, K.R.~Thorne, J.A.~Wheeler.}
    \label{fig:textbook_problem}
\end{figure}

Still in the same approach, planning to attract general attention, on not proved scientific results, this program was inserted in one of the books, expected to become the reference book on Relativistic Astrophysics over shadowing the classical book of Landau and Lifshitz,\cite{Landau1980} a still more extreme proposal of an advanced civilization building a structure around a Kerr BH (see Fig.~\ref{fig:textbook_problem}).

As we are going to show here, the conceptual framework underlying such works missed the understanding of the Black Hole physical laws developed in those years and only recently manifested after $50$~years of considerations.

\subsection{The first fundamental definition: the last stable orbit in a Kerr Black Hole}\label{sec:kerr-bh-last-orbit}
\begin{figure}
    \centering
    \includegraphics[width=1\linewidth]{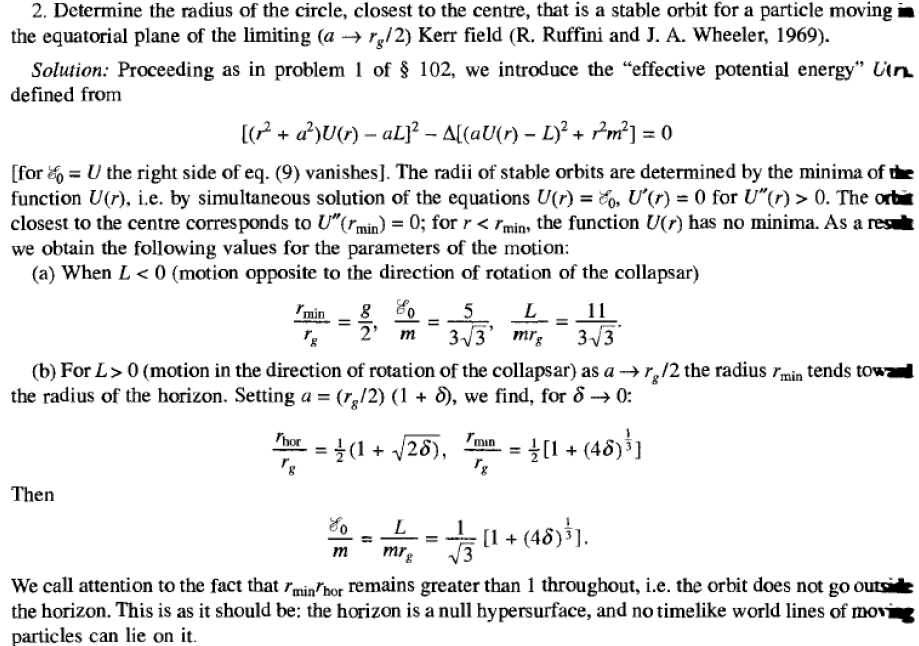}
    \caption{The reproduction of the textbook problem in the ``Theory of fields'' by Landau and Lifshitz (1980).}
    \label{fig:rr-wheeler-textbook-problem}
\end{figure}

Among the first theoretical physics application of the Kerr solution, there was the work of Ruffini and Wheeler (1971)\cite{RR-Wheeler1971},  published in Landau and Lifshitz (1980),\cite{Landau1980} see Fig.~\ref{fig:rr-wheeler-textbook-problem}.

\subsection{The irreducible mass and the ergosphere}\label{sec:irreducible-mass}
\begin{figure}[ht!]
    \centering
    \includegraphics[width=0.7\linewidth]{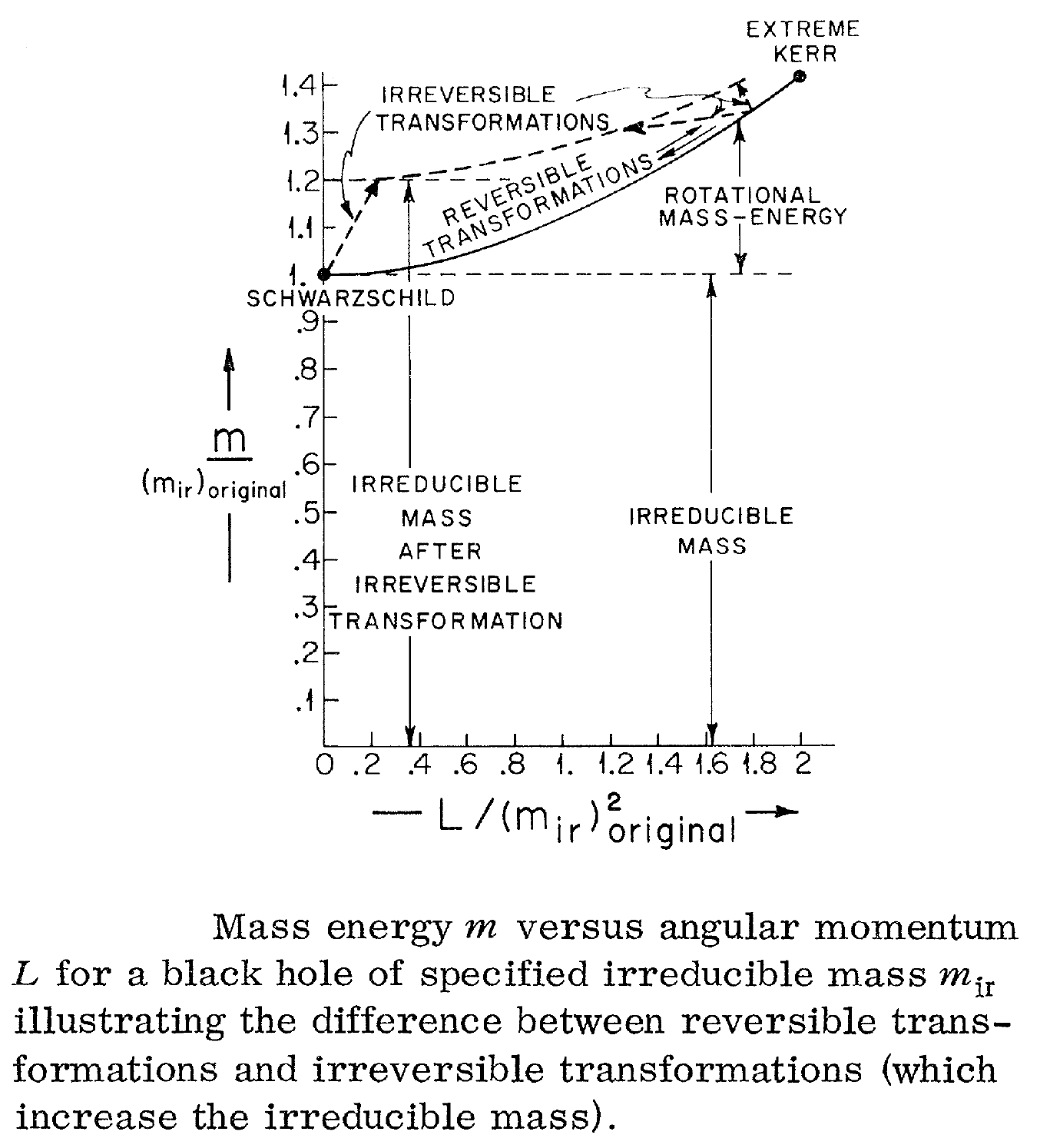}
    \caption{Reproduced from Christodoulou~(1970).}
    \label{fig:christodoulou1970-page1}
\end{figure}

\begin{figure}[hb!]
    \centering
    \includegraphics[width=.8\linewidth]{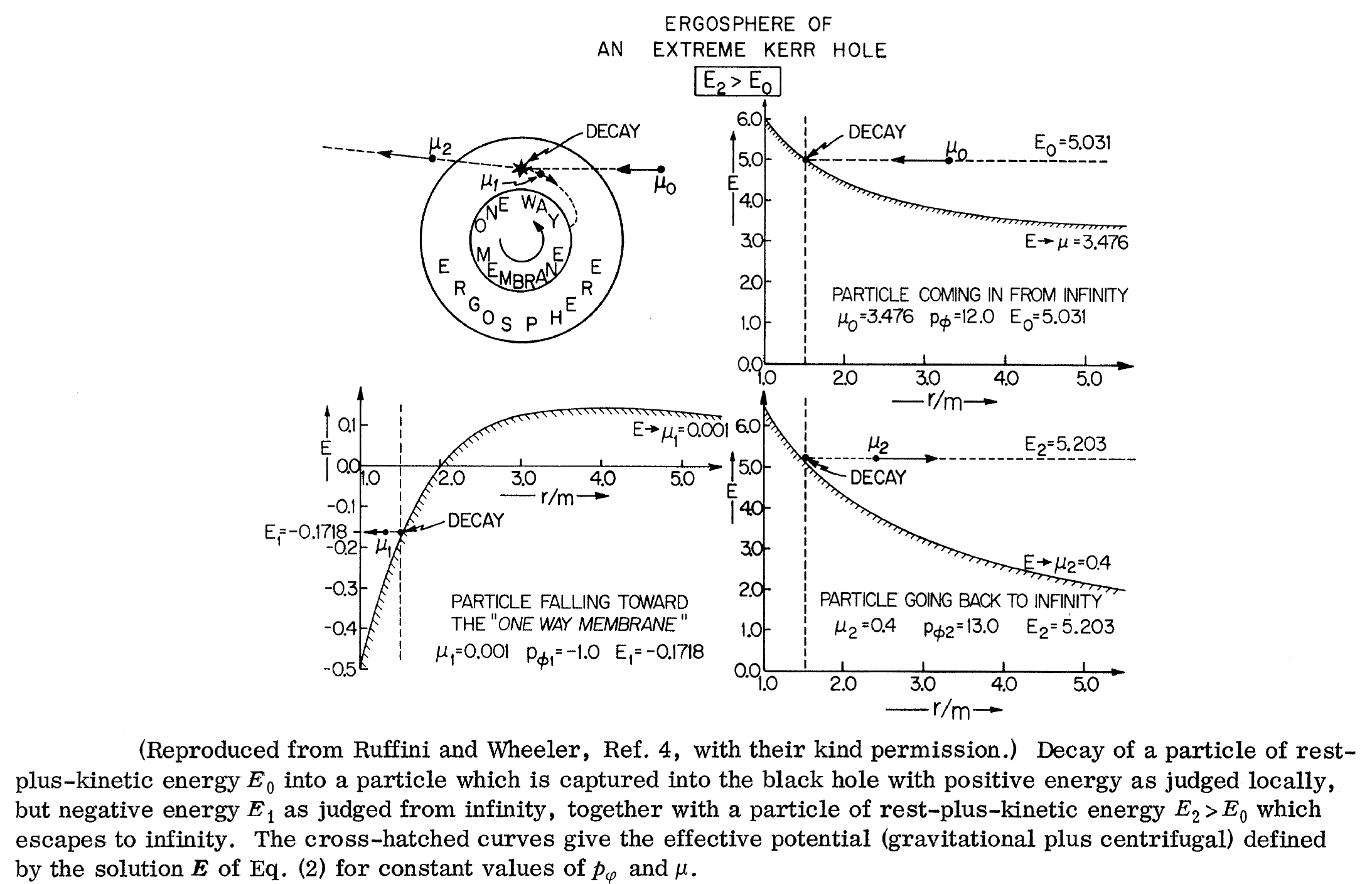}
    \caption{Reproduced from Christodoulou~(1970).}
    \label{fig:christodoulou1970-page2}
\end{figure}

This work was soon followed by the work of Christodoulou~(1970),\cite{Christodoulou1970} see Fig.~\ref{fig:christodoulou1970-page1}. In that paper, the results obtained by Ruffini and Wheeler introducing the concept of Dyadosphere were reported (Fig.~\ref{fig:christodoulou1970-page2}), using the effective potential technique introduced in Landau and Lifshitz, and finally giving the first example of a Penrose process.


\section{The mass-energy formula of a Kerr-Newman Black Hole}\label{sec:mass-energy}
\begin{figure}
    \centering
    \includegraphics[width=1\linewidth]{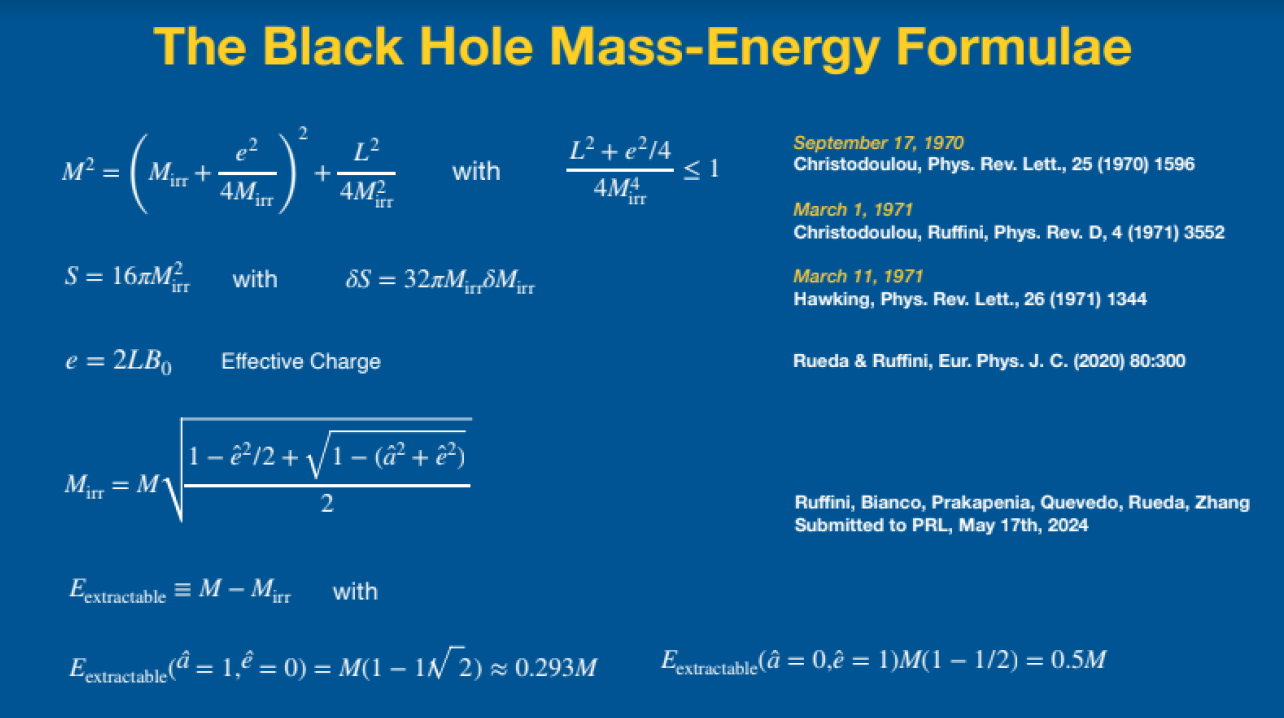}
    \caption{The summary of the BH mass-energy formula as of February, 25th, 2025.}
    \label{fig:bh-mass-energy-formulae}
\end{figure}

\begin{figure}
    \centering
    \includegraphics[width=.8\linewidth]{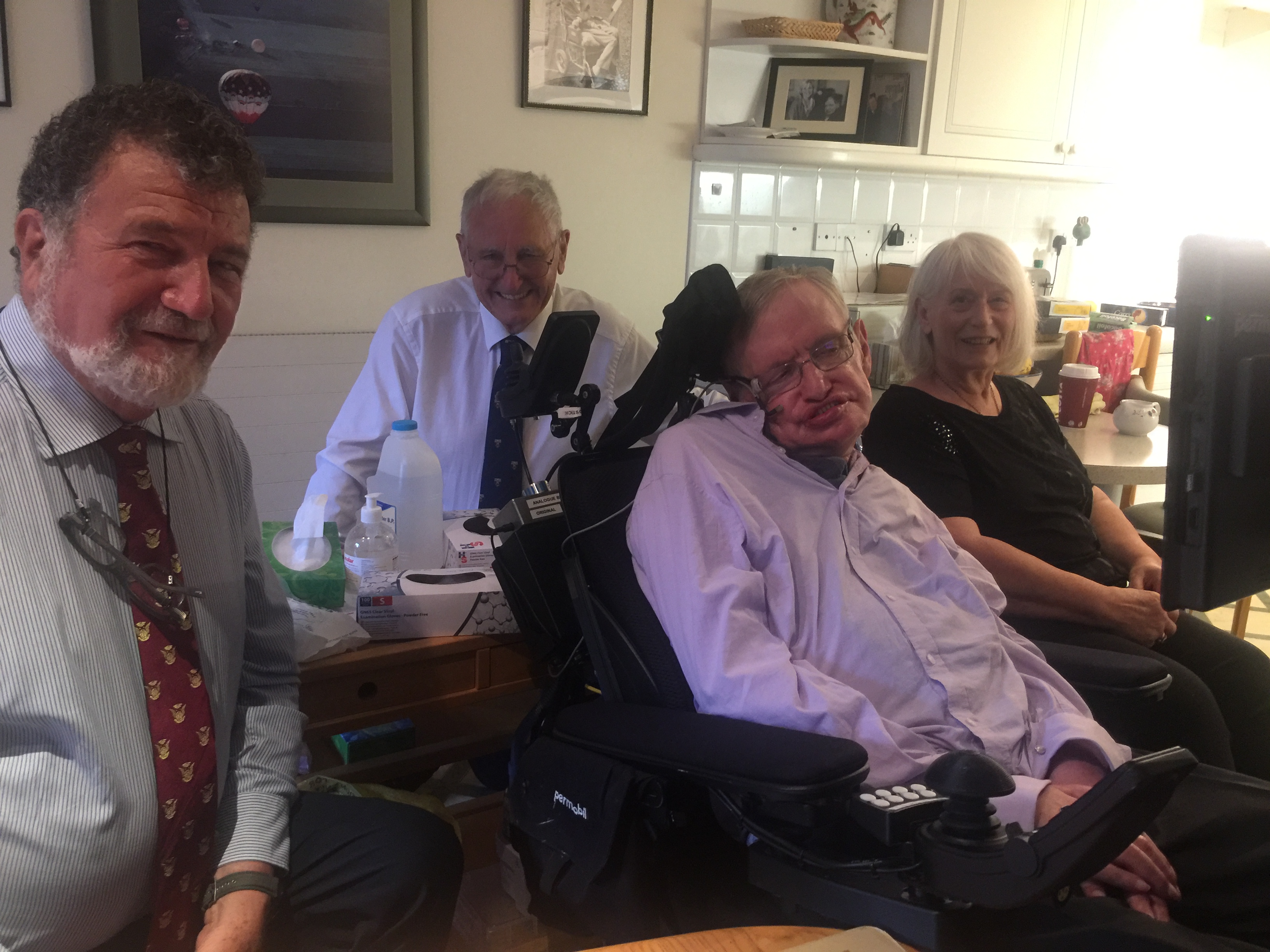}
    \caption{This picture was taken in the Prof. Stephen Hawking's residence in Cambridge at the time which Roy Kerr received at Stockholm the Crafoord Prize. Date: June 20th, 2017.}
    \label{fig:RR-Kerr-Hawking}
\end{figure}

Soon after, a sequence of fundamental works, including Hawking (1971)\cite{Hawking1971} appeared, also the generalized BH mass-energy formula (see Fig.~\ref{fig:bh-mass-energy-formulae}), the relation between the surface area and the $M_{\rm irr}$, the equally fundamental concept of effective charge, the inverse relation between the $M_{\rm irr}$ and the BH parameters as well as the new concepts of extracted energy and extractable energy. In 2017, we had an occasion to discuss the progress done in this field, see Fig. \ref{fig:RR-Kerr-Hawking} – Stephen Hawking, Roy Kerr and Remo Ruffini.

Two papers have since appeared: 
\begin{enumerate}
    \item ICRA-ICRANet press release \textit{``New Study Sheds Light on the Penrose Process and Energy Extraction from Kerr Black Holes''}, March 13, 2025 on the article R.~Ruffini, M.~Prakapenia, H.~Quevedo, and S.~Zhang, \textit{Single versus the Repetitive Penrose Process in a Kerr Black Hole}, Phys. Rev. Lett. 134, 081403 (2025).\cite{PRL2025} Link: \url{https://www.icranet.org/documents/press_release_PRL.pdf}\\
    \item ICRA-ICRANet press release \textit{``New Research Unveils the Role of Irreducible Mass in Energy Extraction from Kerr Black Holes''}, March 13, 2025 on the article R.~Ruffini, C.L.~Bianco, M.~Prakapenia, H.~Quevedo, J.A.~Rueda, and S.~Zhang, \textit{Role of the irreducible mass in repetitive Penrose energy extraction processes in a Kerr black hole}, Phys. Rev. Research 7, 013203 (2025).\cite{RR-PRR2025} Link: \url{https://www.icranet.org/documents/press_release_PRR.pdf}\\
\end{enumerate}

A groundbreaking study published in Physical Review Letters (PRL, 134, 081403, 2025),\cite{PRL2025} explores the long-debated Penrose process, shedding new light on the extraction of rotational energy from Kerr black holes. Led by a team of international researchers from ICRANet and other global institutions, the study provides novel insights into the fundamental physics of energy extraction of astrophysical black holes.

The research, titled ``Single versus the Repetitive Penrose Process in a Kerr Black Hole'' (PRL, 134, 081403, 2025),\cite{PRL2025} revisits the original Penrose process, a theoretical model proposed by Roger Penrose in 1969.\cite{Penrose1969} The study examines how a single decay event of a massive particle into two particles inside the ergosphere of a rotating black hole can result in energy extraction. The team successfully demonstrated that, contrary to earlier criticisms, the single-event Penrose process is indeed capable of extracting significant energy efficiently, with an upper efficiency of $14.5\%$ for a maximally rotating Kerr black hole.

A key aspect of the study also explores the possibility of a repetitive Penrose process, first suggested by Misner, Thorne, and Wheeler in ``Gravitation'' (1973), see Fig. \ref{fig:textbook_problem}. The researchers theoretically analyzed an iterative sequence of decay processes, which, if naively implemented, would appear to extract $100\%$ of the rotational energy of a BH. However, their findings highlight a crucial limitation: such a linear repetitive process would violate energy conservation laws. The team highlights that this inconsistency can  be solved by properly incorporating the nonlinearity introduced by the increase of the irreducible mass of the black hole during the process, which is the subject of an accompanying publication in Physical Review Research. These results laid the groundwork for a revised energy extraction mechanism that obeys fundamental conservation principles.

Our study, instead, clarifies the feasibility of the Penrose process as a mechanism for energy extraction from Kerr black holes. We have demonstrated that a single Penrose process is indeed effective, no modifications are needed  as  previously assumed. The authors  have also uncovered the key limitations of a naive repetitive approach and, in the specific examples provided, evidenced significant  mass defect constraints. This research provides a necessary step toward a more complete understanding of black hole energetics.

These findings have broad implications for astrophysics, including the study of high-energy cosmic phenomena such as gamma-ray bursts and active galactic nuclei, where black hole rotational energy may play a critical role. Future research will aim to expand on these results, further refining our understanding of black hole dynamics and potential astrophysical applications.

The accompanying paper published in Physical Review Research (7, 013203, 2025)\cite{RR-PRR2025} was conducted by researchers from ICRANet, ICRA, INAF, Al-Farabi Kazakh National University, the University of Science and Technology of China, the Universidad Nacional Aut\'onoma de M\'exico, and other institutions. The work builds on decades of theoretical advancements in BH physics, combining classical general relativity with modern astrophysical insights.


A new study offers a major breakthrough in our understanding of the Penrose process, refining our knowledge of energy extraction from Kerr black holes. Building on a previous study in PRL, this work clarifies the limitations of the repetitive Penrose process and establishes the central role of irreducible mass in black hole dynamics, see Fig. \ref{fig:christodoulou1970-page2} and Fig. \ref{fig:bh-mass-energy-formulae}.

The study, titled the ``Role of the Irreducible Mass in Repetitive Penrose Energy Extraction Processes in a Kerr Black Hole'', highlights the highly nonlinear nature of energy extraction from a rotating black hole. The research team, led by scientists from ICRANet and global institutions, demonstrates that the increase in a black hole’s irreducible mass significantly limits the efficiency of repetitive Penrose processes, debunking prior assumptions about the feasibility of extracting the entirety of a black hole’s rotational energy.

Our findings show that, contrarily to previous expectations, the Penrose process is far from a linear, scale-invariant mechanism. Instead, the irreducible mass increases much more than the extracted energy, ultimately halting the process after a finite number of iterations.

Key results of the study include:\\
\begin{enumerate}
    \item Nonlinear limitations: The research demonstrates that each successive decay event in the ergosphere of a Kerr BH leads to an increase in irreducible mass, which quickly outweighs the extracted energy.\\
    \item Finite energy extraction: The iterative Penrose process stops well before a black hole can be fully stripped of its rotational energy. The latter amounts to a maximum of 29\% of the BH mass if it is at maximal rotation. For selected decay radii inside the BH’s ergosphere and a decaying particle mass of 1\% of the mass of a maximally rotating BH, the paper shows that energy extraction ceases after as few as 8 or as many as 34 iterations, extracting at most 1\% and 0.4\% of the BH’s mass, while reducing the rotational energy by 17\% and 50\%.\\
    \item Breakdown of naive repetition models: The above implies that the work corrects prior assumptions that a sufficiently large number of iterations could extract 100\% of a BH’s rotational energy. Instead, the process is self-limiting, as rotational energy in the process is primarily converted into BH’s irreducible mass rather than released to infinity.\\
\end{enumerate}

An example of repetitive Penrose process in a maximally rotating black hole examined in the paper, can be found in the video snapshots of the repetitive Penrose process (Fig.~\ref{fig:snapshot001}):

\begin{figure}
    \centering
    \includegraphics[width=\linewidth]{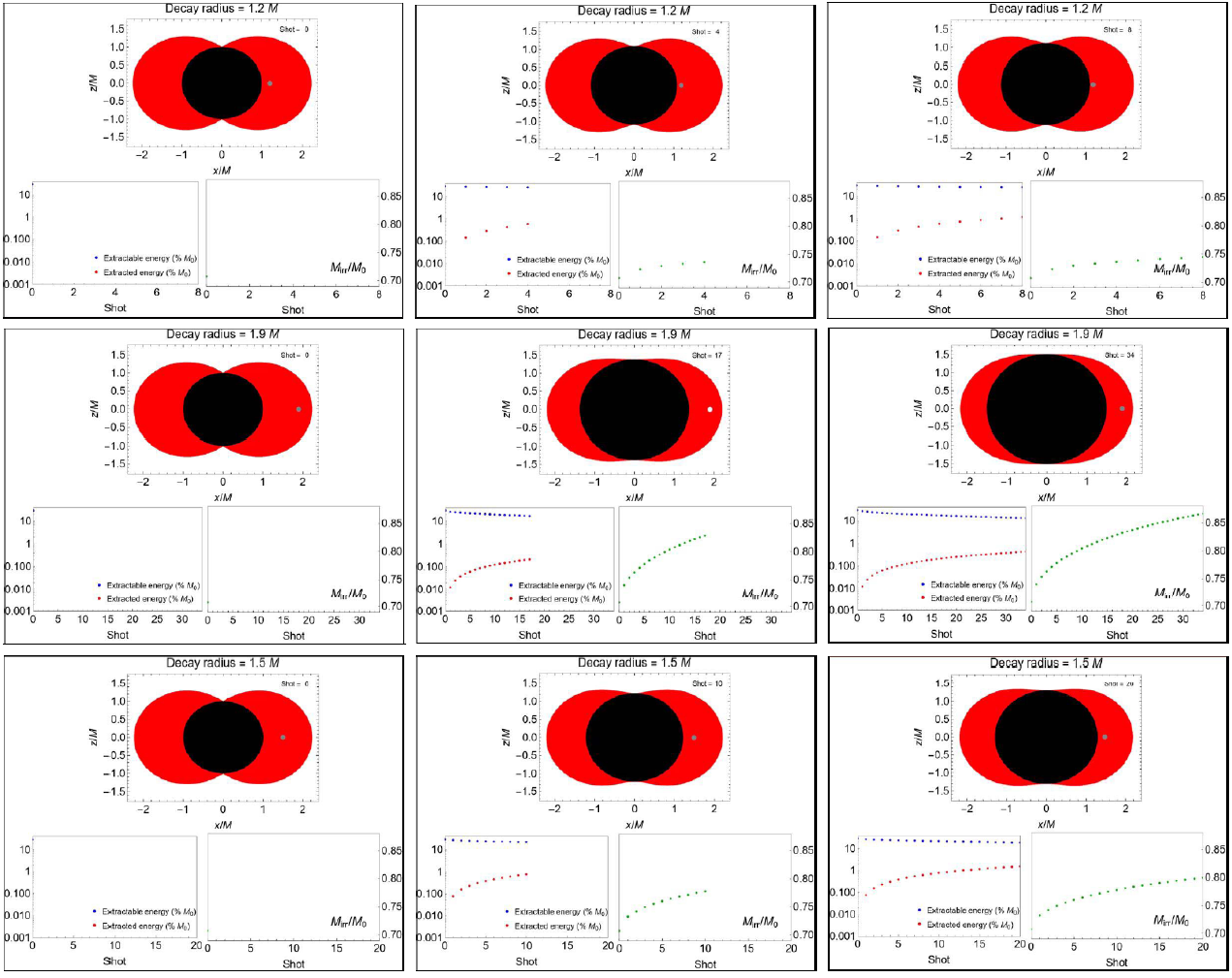}
       \caption{Snapshot 0, Snapshot 4, Snapshot 8, see \url{www.icranet.org/documents/video_dot1.mp4}.
       Snapshot 0, Snapshot 17, Snapshot 34, see \url{www.icranet.org/documents/video_dot2.mp4}.
       Snapshot 0, Snapshot 10, Snapshot 20, see \url{www.icranet.org/documents/video_dot3.mp4}.}
    \label{fig:snapshot001}
\end{figure}

These findings have profound implications for astrophysical models involving BH energy extraction well identified in gamma-ray bursts and other high-energy cosmic events. 

Together, the above two studies provide a comprehensive revision of the Penrose process and its constraints, implying the need for alternative processes to extract the BH's energy keeping the irreducible mass increase as small as possible. In this line, electrodynamical processes appear as a promising path as shown by recent works of the team.

The publication of the original results goes back fifty years and has been followed for many decades of thinking and progress reaching the final form in these days:  it is fortunate that the main contributors: Roger Penrose and Remo Ruffini have enjoyed sharing and discussing these results and indicate further developments (see Enclosure 1 \textit{``Roger Penrose to Remo Ruffini, February 23, 2025''} and Enclosure 2 \textit{``Remo Ruffini to Roger Penrose, February 26, 2025''}, available in the above press release \url{https://www.icranet.org/documents/press_release_PRL.pdf}).

An essential  point to address is the objective scientific urgency of the following results: since the first discovery of a BH in Cygnus X-1 in a binary X-ray source (Ruffini R., N.Y. Texas Meeting 1972), based on the absolute upper limit on the NS $M_{crit}=3.2M_{\rm odot}$, see Rhoades and Ruffini in PRL (32, 324, 1974).\cite{Rhoades-RR_1974} The largest observational effort in developing observatories from the ground and from space have reached the scope ``Identify  BHs all over the Universe'': from inside our galaxy to extragalactic sources all the way to the highest cosmological redshift at $z=10$ and higher prior to the decoupling era. The BHs, typically of $2M_{\odot}$--$10M_{\odot}$, originate from the collapse of a ``baryonic matter''. The great novelty is the current discovery of BHs much larger masses, at $4.6\times10^6M_{\odot}$ all the way to $10^{10}M_{\odot}$ originating from Dark Matter and manifesting themselves from cosmological $z=10$ in supermassive BHs, all the way to $z=2$ under the form of Quasars.

Precisely this topic has been indicated as a high priority in the Penrose-Ruffini exchange. Progress has been made in ICRANet by Arguelles, Rueda, Ruffini (January 11, 2024); Ruffini-Vereshchagin March 2025, and Ruffini, Della Valle-Wang Yu (March 2025). They examine BH composed of both Dark Matter and Baryonic matter and determine their evolution. Of great relevance for this new paradigm are the observations of the, so-called, Little Red Dots by the James Webb Space Telescope of NASA: ``Red-Dots messenger of  CDMB, a Cosmic Dark-Matter Background'' coeval to the better known ``CMB Cosmic Microwave Background''.

\section{Some conclusions}
Following the above results have been: \textit{``The evaluation of the increase of the irreducible mass in any energy extractable process in the Kerr BH ergosphere is mandatory''}.

In order to evaluate the amount of the final extracted energy from rotating Kerr BH, it is necessary to ascertain the reversibility or the irreversibility of the process and, correspondingly, to establish the extractable energy.\cite{Rueda-RREPJC2023}

If one addresses solely gravitational interactions, the formulae given above show a dependence on the $M_{\rm irr}$ which cannot be neglected. Both in the case of single and repetitive extraction processes, the extracted energy is smaller than $3\%$ of the initial BH rotating mass. What was unexpected is that the (final!) result of a repetitive process leads to the extraction of the entire rotational energy and the entire mass energy of the initial BH consists only of the irreducible mass.

Markedly different is the case of a Kerr-Newman BH, presence of a background field, endowed with an effective charge in which the rotational energy extraction process approaches reversibility \cite{Rueda-RR-KerrApJ2022}
\cite{Rueda-RREPJC2023}
\cite{Rueda-RREPJC2024}.


\section*{\Large The most energetic BdHN I}

\section*{Introduction}\label{sec:part2-introduction}
\begin{figure}
    \centering
    \includegraphics[width=\linewidth]{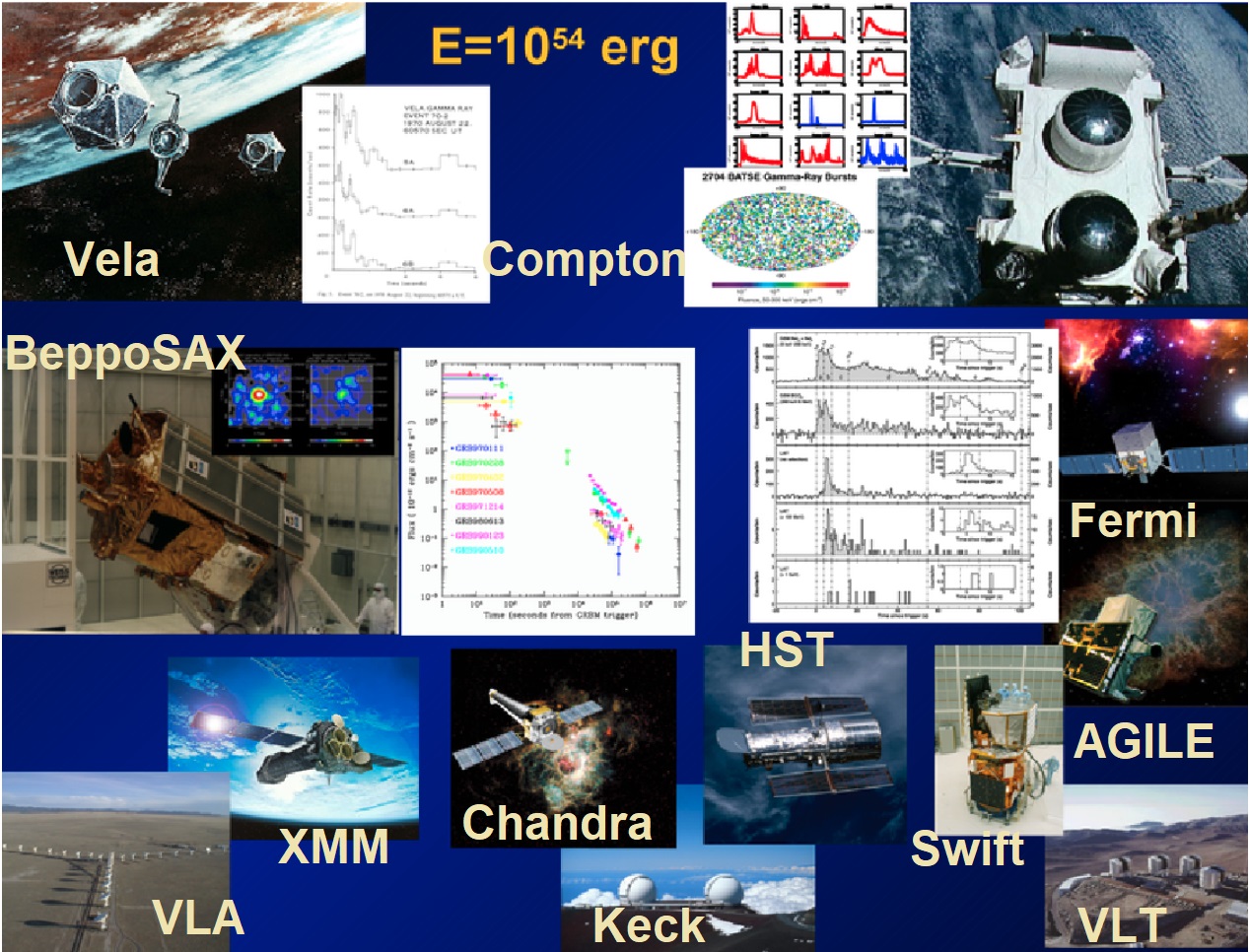}
    \caption{GRB dedicated space missions.}
    \label{fig:missions}
\end{figure}

The physics and astrophysics of GRBs was developed over a time span of more than fifty years, with a very central role played by the BeppoSax telescope, well summarized in Fig.~\ref{fig:missions}.

Here, the moment of discovery by the Vela satellites, publicly announced at the AAA meeting in San Francisco, summarized in the book by Gursky and Ruffini (Neutron Stars, Black Holes, and Binary X-ray sources, D. Reidel Publishing Company, 1975),\cite{Gursky-RR_1975} was then followed by the discovery of the Compton satellite working in the Gamma ray and creating the first distribution of GRBs in galactic coordinates and showing clear homogeneous distribution, independently of the energetics (represented by the colored dots in the image), which gave the first clear evidence for the extragalactic nature of the GRBs, as well as their clear evolution on time scales of the order of milliseconds, but it was not. At the time, no knowledge of their distance had been known and, therefore, of their luminosity. 

It was the intervention of the X-ray telescopes, introduced by BeppoSAX and their narrow field instruments, that the discovery of the X-ray counterpart of the Gamma Ray observations, that allowed the observation of the X-ray afterglow and, consequently, the localization sufficiently accurate to allow their optical identification and, from the optical redshift, the determination of their cosmological distances and enormous energetics. It was a fortunate coincidence that, precisely in the same year, the largest optical telescope on Mauna Kea, the Keck Telescopes, became operative and the development of a new class of telescopes in all wavelengths became possible. 

The next fundamental observational breakthrough occurred with the Fermi Telescope operative in the GeV radiation which, as we will see shortly, allowed the first identification of BH in GRB, allowed to estimate the GeV afterglow in a selected GRBs, see Ruffini et al. MNRAS (2021)\cite{RR-MNRAS_2021}, allowed to determine the beaming angle of the radiation as the lower limit to the GeV emission. In order to establish a more precise value of the Kerr BH mass we had to wait for the results of Roger Romani and colleagues (2022)\cite{Romani_2022} on the absolute upper limit to the observed NS in GRB 220101A in 2025.

\begin{figure}
    \centering
    \includegraphics[width=\linewidth]{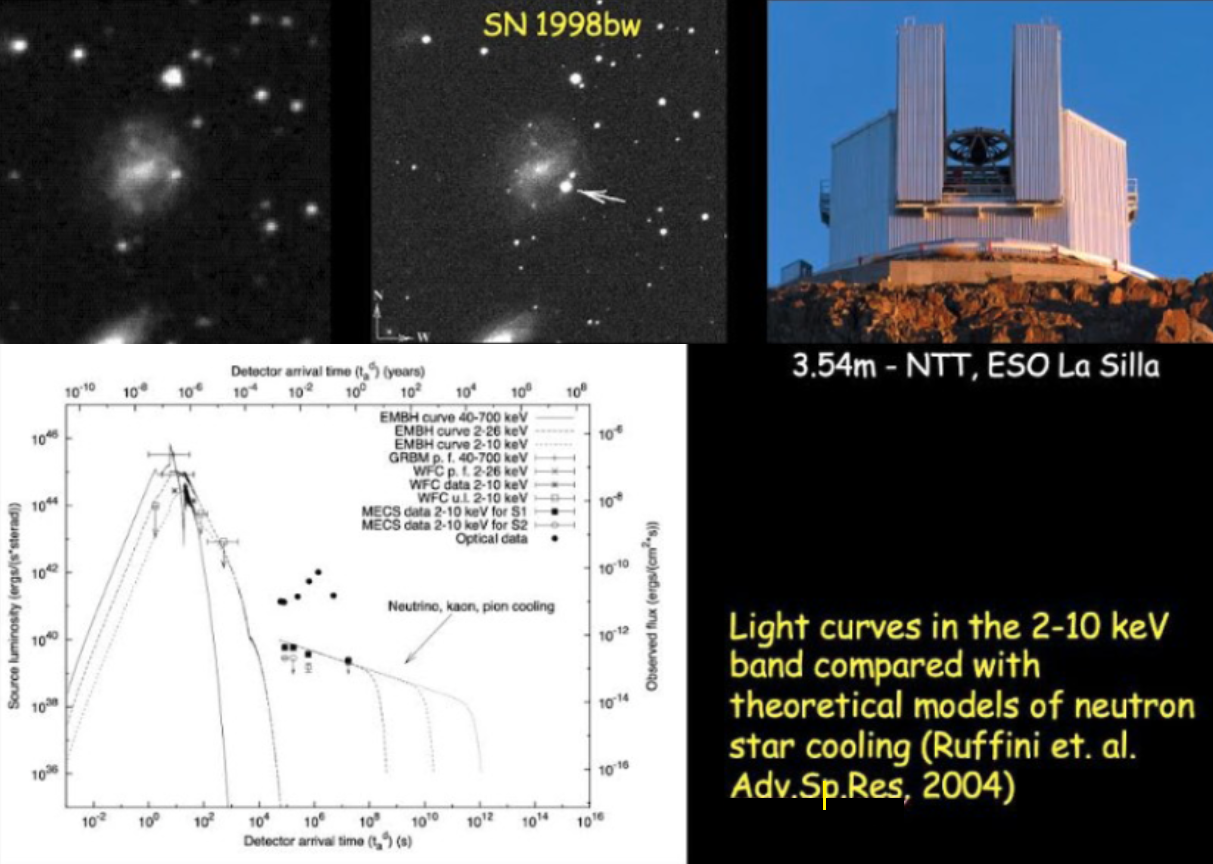}
    \caption{Supernova SN~1998bw associated with GRB.}
    \label{fig:sn1998bw}
\end{figure}

\begin{figure}
    \centering
    \includegraphics[width=\linewidth]{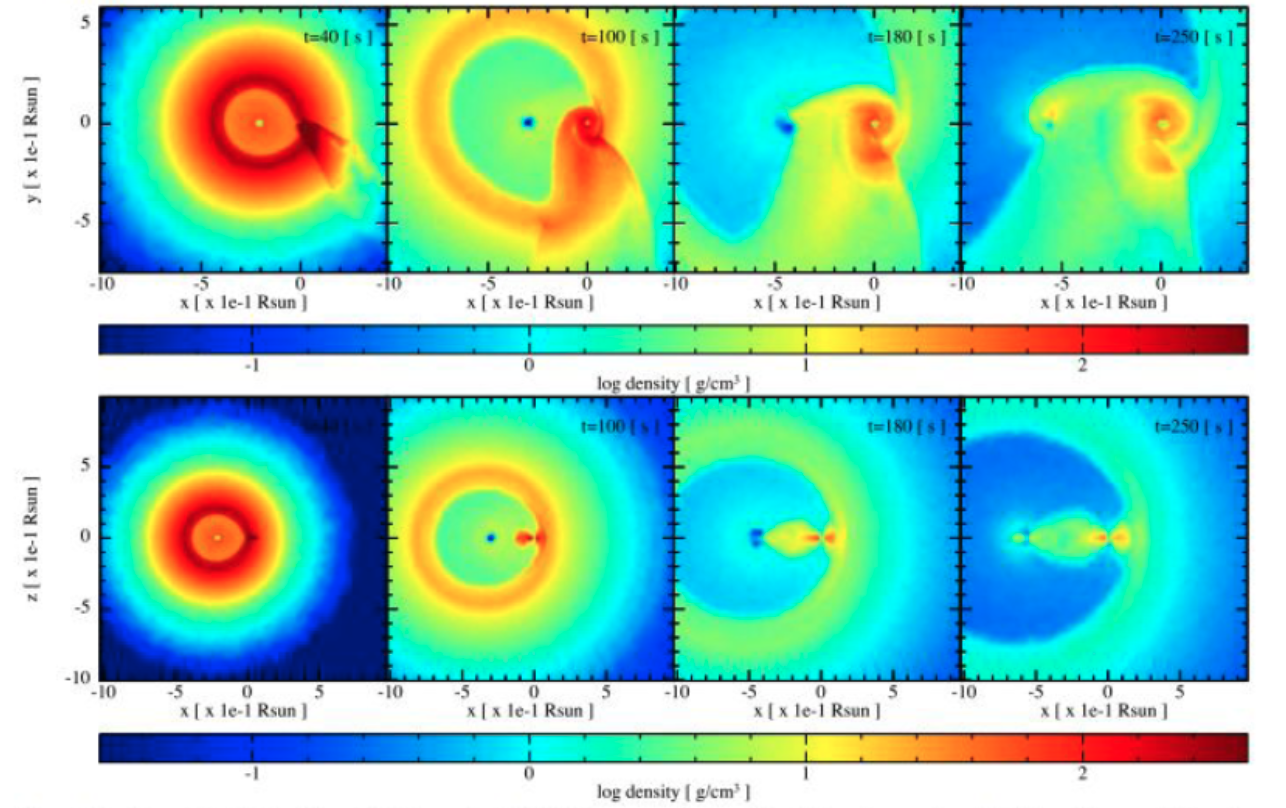}
    \caption{Snapshots of the SPH simulation of the Induced Gravitational Collapse (IGC) scenario. The initial binary system is formed by a CO-core, the progenitor of which is an $M_{\rm ZAMS} = 25M_{\odot}$, and a $2M_{\odot}$ NS with an initial orbital period of approximately $5$~minutes. The upper panel shows the mass density on the binary equatorial plane at different times of the simulation, while the lower panel corresponds to the plane orthogonal to the binary equatorial plane. The reference system was rotated and translated in such a way that the x-axis is along the line that joins the binary stars and the origin of the reference system is at the NS position. At $t = 40$~s (first frame from left), it can be seen that the particles captured by the NS have formed a kind of tail behind it, then these particles star to circularize around the NS and a kind of thick disk is observed at $t = 100$~s (second frame from left). The material captured by the gravitational field of the NS companion is also attracted by the $\nu$NS and starts to be accreted by it, as can be seen at $t = 180$~s (third frame). After around one initial orbital period, at $t = 250$~s, a kind of disk structure has been formed around both stars. The $\nu$NS is along the x-axis at $-2.02$, $-2.92$, $-3.73$, and $-5.64$ for $t = 40$, $100$, $180$, and $250$~s, respectively. Note that this figure and all snapshot figures were done with the SPLASH visualization program.}
    \label{fig:sph}
\end{figure}

One of the crucial discovery to stimulate the connection between a GRB and a Supernova was the temporal and spatial coincident position of a GRB~980425 and  a SN~1998b, so profoundly different in their spectral distribution and in their energetics, see details in Fig.~\ref{fig:sn1998bw} and Ruffini et al. Advances in Space Research (2004).\cite{RR-COSPAR_2004} This was certainly one of the leading discoveries which promoted our change of paradigm, to move from a GRB originating from a single object, a collapsar, to a binary system composed of a binary system composed of a CO-core of $10M_{\odot}$ and a binary NS companion in a system with a period of a few minutes. This was the basis of the Binary driven hypernova (BdHN), further examined by simulation developed in Los Alamos National Laboratory, see Fig.~\ref{fig:sph}.

\begin{figure}
    \centering
    \includegraphics[width=0.8\linewidth]{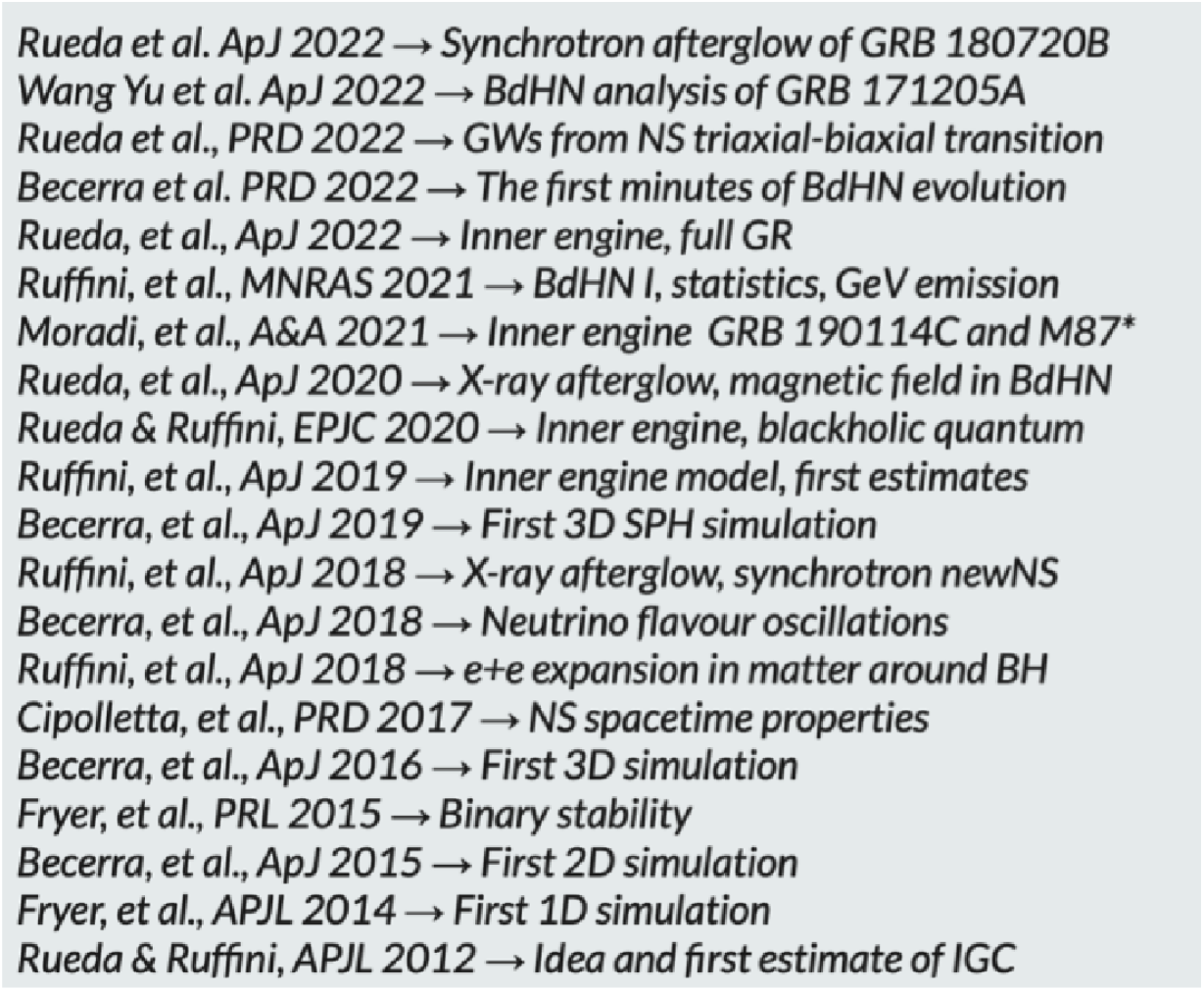}
    \caption{Chronology of the theoretical development of the Binary-driven Hypernova model and its key aspects published in a series of articles.}
    \label{fig:chronology}
\end{figure}

Some of the developments of the key aspects can be found in the following references (see Fig.~\ref{fig:chronology}).

\section{The summary of some of our fundamental contribution originating from GRB observations}\label{sec:part2-summary-fund-contrib-grb}

Our GRB analysis is still daily evolving following some conceptual changes which resulted from dialogues that have occurred through fifty years. I have recalled the conceptual black hole electrodynamics of collaboration in Princeton with John Wheeler and various students including Demetrios Chrsitodoulou and other Princeton students. 

An intense data analysis which occurred practically everyday following the Galileo's principle \textit{``provando e riprovando''} since the daily scrutiny of the BeppoSAX satellite started and exponentially expanded in the following decades extending to the largest ever observational multi-wavelength effort in the history of science which slowly and constantly evolved the GRB model to the final, reaching the success of GRB~220101A.

There had been also personal commitment  which significantly influenced our work: I had promised Werner Heisenberg in our first meeting in Munich in the early seventies to have Black holes to become a test observational field for proving the occurrence of the Euler-Heisenberg process and pair creation, see, e.g., R. Ruffini ``Einstein, Fermi, Heisenberg and the birth of relativistic astrophysics'', World Scientific, (2025, in press). They were submitted to an intense but yet unsuccessful search in Earth-bound observatories both in the US at Brookhaven National Laboratory and in Germany. This motivated us to develop a methodic analysis of assuming the presence of an electrodynamical structure in a BH, leading to the first example of a process of vacuum polarization around an overcritical electrodynamical field in Damour and Ruffini (1975).\cite{Damour-RR_1975} Various processes of vacuum polarization led to the discovery of the dynamics of a spherical symmetric $e^+e^-$ plasma self accelerating in a pair-electromagnetic (PEM) pulse were introduced in a collaboration with Jim Wilson in 1999.\cite{RR-Salmonson-Wilson-Xue_AandA_1999,RR-Salmonson-Wilson-Xue_AandASupl_1999} Soon after the fundamental concept of a PEMB pulse composed of pair and baryons was introduced also with Wilson.\cite{RR-Salmonson-Wilson-Xue_AandA_2000} An important progress in understanding the origin of an electrodynamical structure came from the work of Papapetrou, examining the properties of a Kerr BH, not in vacuum but in presence of a background magnetic test field also discussed by Robert Wald.\cite{Wald_1974} This problem was brought to an operative astrophysical system by Rueda and Ruffini.\cite{Rueda-RR_2012,Rueda-RR_2020} These were important conceptual developments which preceded, accompanied and follow our understanding of astrophysical conditions in GRBs and now are present in a much larger scenario involving not only baryonic matter Black Holes but also dark matter.

Without going too far back in time I am going to consider a few examples which have necessarily lead to a change of our paradigms in relativistic astrophysics being guided by the astrophysics of GRBs. Offering a vast statistical distribution of the SNe associated to GRBs,\cite{Aimuratov_2023} and also the first clear manifestation of seven characteristic Episodes in the most detailed GRB, we have conducted to the first the analysis of GRB~220101A. It is from the details gained in some specific sources that an unprecedented progress is just starting. While the seven episodes have gained their visibility and now are generally accepted, a new wave is just starting which we only briefly address and summarized in Ruffini, Mirtorabi et al. (2025, in~press)\cite{RR-Mirtorabi2025}, where seven episodes were used as a tool to learn about the structure of SN needed from observation: the triggering due to a pair SN followed a few seconds after by a more traditional core-collapse SN. 

%
%

\section{The first evidence for reaching overcritical electrodynamical field occurred in the GRB~190114C}\label{sec:part2-overcritical-field-190114c}

\subsection{GRB~190114C and Ultra-relativistic Prompt Emission (UPE) phase}\label{sec:part2-190114c-upe-phase}

\begin{figure}
    \centering
    \includegraphics[width=\linewidth]{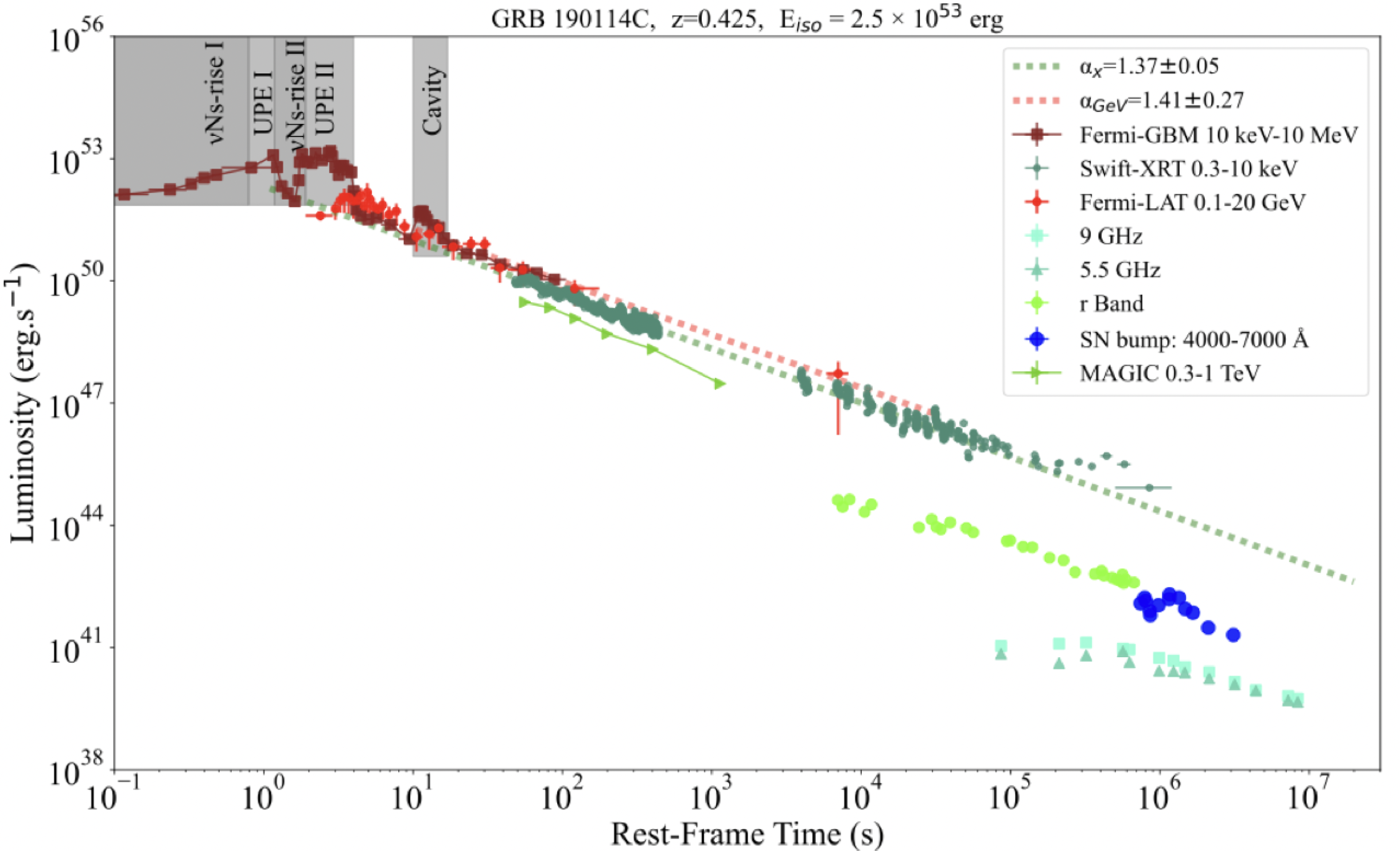}
    \caption{Light curve of GRB~190114C obtained from data from Fermi-GBM, Fermi-LAT, Swift-XRT, MAGIC, optical and radio telescopes. Episodes of the BdHN burst evolution are denoted by shaded regions. Reproduced from Aimuratov et al. (2023).}
    \label{fig:grb190114c_lc}
\end{figure}

In the source GRB~190114C, which has so far been the most important, the fundamental GRB has opened an entirely new field of research and comprehension, including the discovery of UPE~I and UPE~II episodes, with an estimated isotropic emission up to $\sim10^{54}$~erg, to identify for the first time the optical counterpart of the SN, to have identified the TeV emission by observation of MAGIC, to have identified GeV emission by observation of Fermi-LAT, see Fig.~\ref{fig:grb190114c_lc}. In addition, the Swift-XRT data have allowed us to identify the X-ray afterglow. These results have been of paramount importance in identifying the three different afterglows, which have been found in all subsequent GRBs, respectively in X-ray, optical, and GeV energy ranges.\cite{Moradi_PhRvD_2021} For the first time, we have extrapolated these light curves to a thousand years and compared them to the spectra of the SN in the Crab Nebula, see Figure~1 in Ruffini and Sigismondi (2024).\cite{RR-Sigismondi2024}

\subsection{The fractal structure of the UPE from GRB~190114C}\label{sec:part2-fractal-structure-190114c}

\begin{figure}
    \centering
    \includegraphics[width=\linewidth]{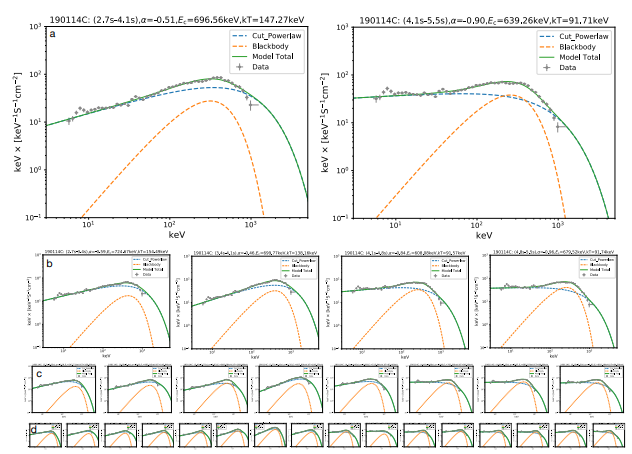}
    \caption{Time-resolved spectral analysis of UPE phase GRB 190114C: from t = 2.7 s (trf = 1.9 s) to t = 5.5 s (trf = 3.9 s).
For the second iteration; (a) the time interval is divided into two parts, four parts for the third iteration; (b), eight parts for
the fourth iteration; (c), and sixteen parts for the fifth iteration; (d), respectively. The spectral fitting parameters for each
iteration are reported in Table I. Plots are taken from Ruffini et al. (2021).}
    \label{fig:grb190114c}
\end{figure}

\subsection{Nature of UPE phase in GRB~190114C}\label{sec:part2-upe-nature-190114c}

We address the physical origin of the ultrarelativistic prompt emission (UPE) phase of GRB 190114C observed in the interval $t_{\rm rf}=1.9$--$3.99$~s, by the {Fermi}-GBM in 10 KeV--10 MeV energy band. Thanks to the high signal–to–noise ratio of {Fermi}-GBM data, {a time-resolved spectral analysis has evidenced a sequence of similar blackbody plus cutoff power-law spectra (BB+CPL), on ever decreasing time intervals during the entire UPE phase.}, see Fig. \ref{fig:grb190114c}. We assume that  during  the UPE phase, the ``\emph{inner engine}'' of the GRB, {composed of a Kerr black hole (BH) and a uniform test magnetic field $B_0$, aligned with the BH rotation axis,} operates in an overcritical field  $|{\bf E}|\geq E_c$, where $E_c=m_e^2 c^3/(e\hbar)$, being $m_e$ and $-e$ the mass and charge of the electron. {We infer an $e^+~e^-$ pair electromagnetic plasma in presence of a baryon load, \emph{a PEMB pulse},  originating from a vacuum polarization quantum process in the {inner engine}.} This initially optically thick plasma self-accelerates, giving rise at the transparency radius to the MeV radiation observed by {Fermi}-GBM. At times $t_{\rm rf}>3.99$~s, the electric field becomes undercritical, $|{\bf E}|<E_c$, and the {inner engine}, {as previously demonstrated}, operates in the  classical electrodynamics regime and generate the GeV emission. { During both the ``quantum'' and the ``classical'' electrodynamics processes, we determine the time varying mass and spin of the Kerr BH in the {inner engine},} fulfilling the Christodoulou-Hawking-Ruffini mass-energy formula of a Kerr BH.  {For the first time, we quantitatively show how the {inner engine}, by extracting the rotational energy of the Kerr BH, produces a series of PEMB pulses. We follow the quantum vacuum polarization process in sequences with decreasing time bins. We compute the Lorentz factors, the baryon loads and the radii at transparency, as well as the value  of the magnetic field, $B_0$, assumed to be constant in each sequence. The fundamental hierarchical structure, linking the quantum electrodynamics regime to the classical electrodynamics regime, is characterized by the emission of ``{blackholic quanta}'' with a} timescale $\tau \sim 10^{-9}$~s, and energy $\mathcal{E} \sim 10^{45}$~erg.

\section{The first systematic examination of the supernova associated with GRB}\label{sec:part2-sn-grb-association}

\begin{figure}
    \centering
    \includegraphics[width=.45\linewidth]{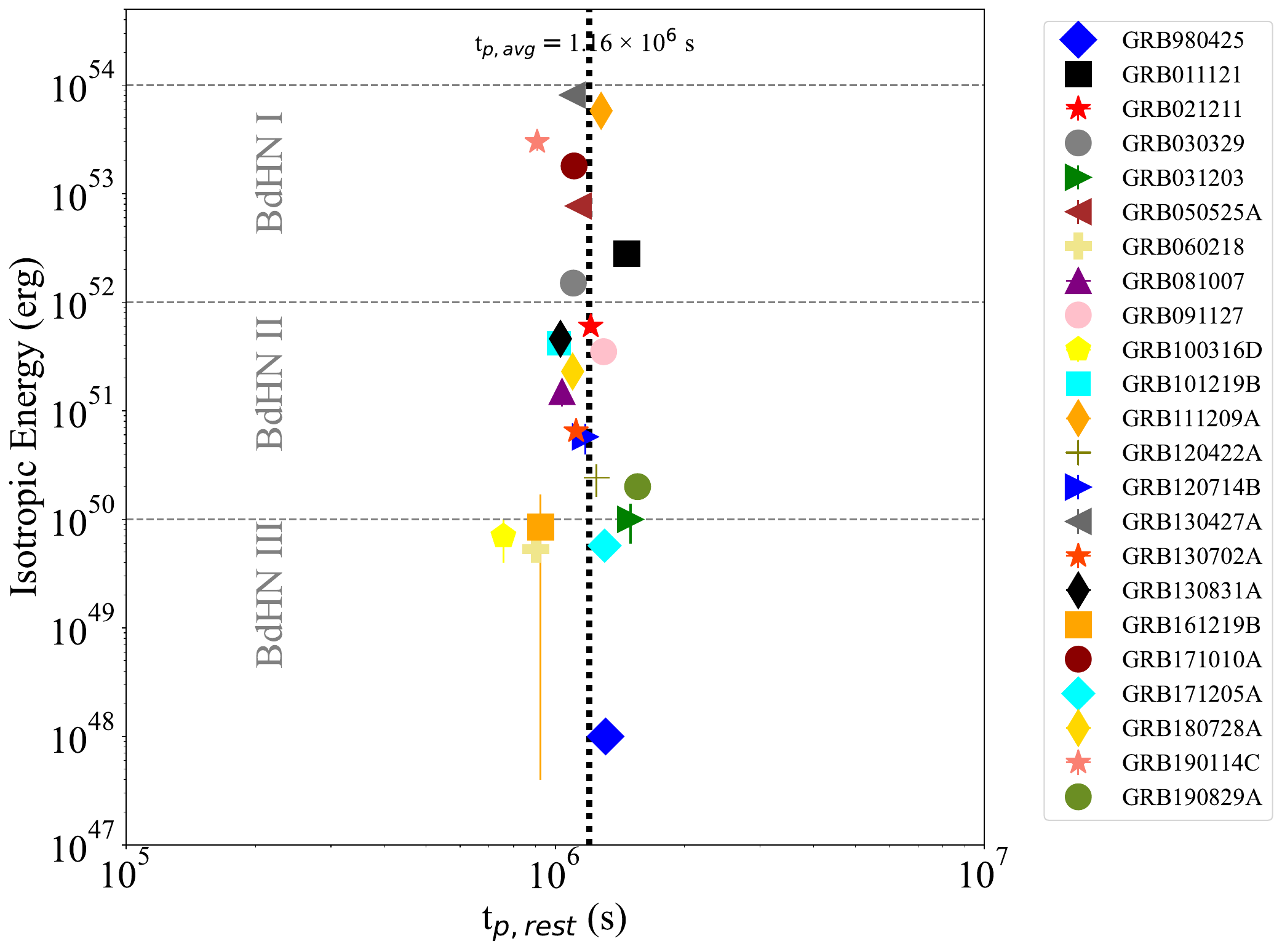}
    \includegraphics[width=.45\linewidth]{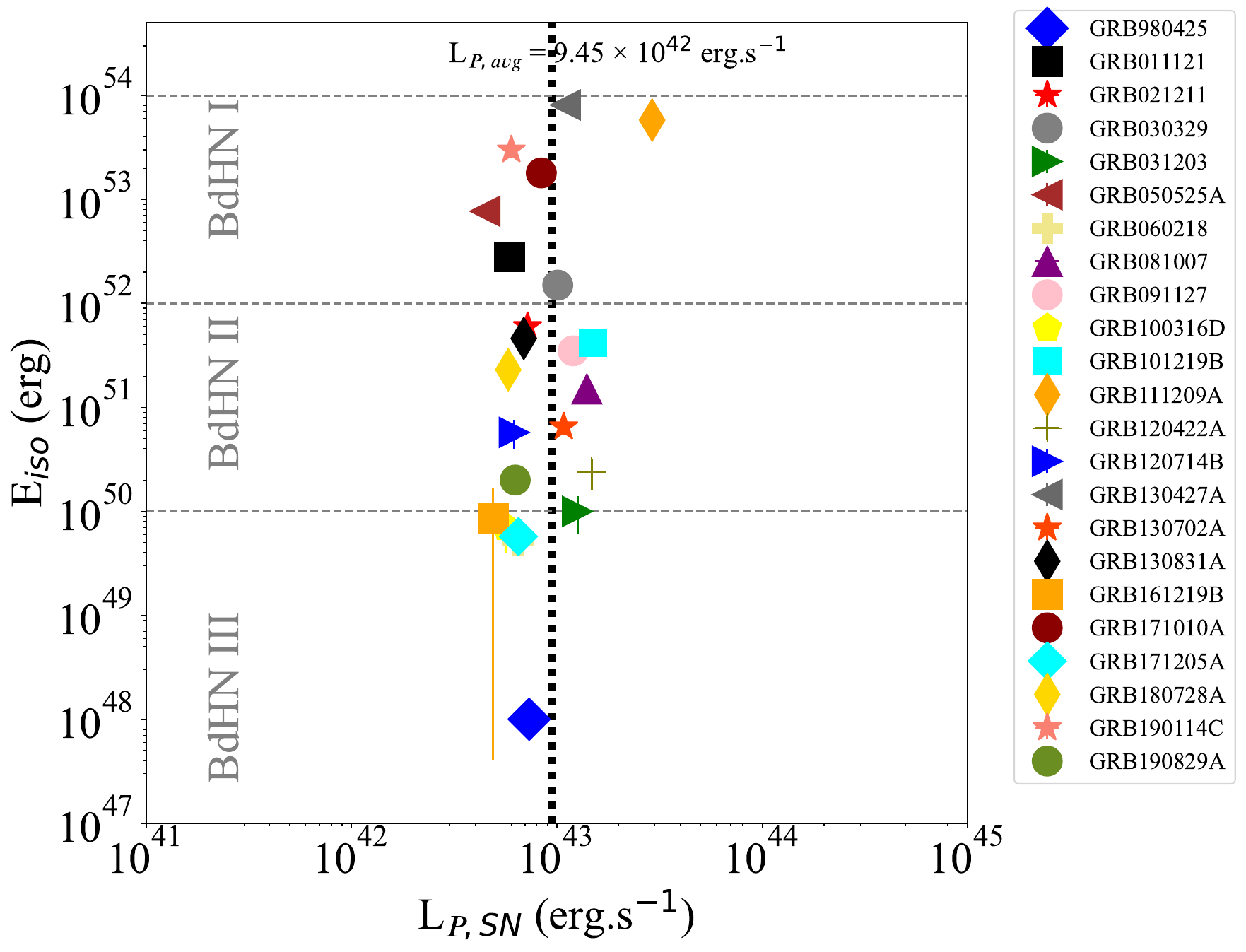}
    \caption{Left panel: Isotropic-equivalent energy ($E_{\rm \gamma, iso}$) of GRB versus the peak time of luminosity of the bolometric light curve of the associated SN ($t_{\rm P, SN}$). The plot shows the lack of correlation: the SN peaking times (in the rest-frame) stay within an order of magnitude spread, while the GRB energy spans $\sim$6 orders of magnitude. Right panel: Isotropic-equivalent energy ($E_{\rm \gamma, iso}$) of GRB versus the peak luminosity of the bolometric light curve of the associated SN ($L_{\rm P, SN}$). The plot shows the lack of correlation: the SN luminosities stay within an order of magnitude spread, while the GRB energy spans $\sim 6$ orders of magnitude. Plots are taken from Ruffini et al. (2023).}
    \label{fig:Eiso-Tpeak-SN}
\end{figure}

\begin{figure}
    \centering
    \includegraphics[width=.45\linewidth]{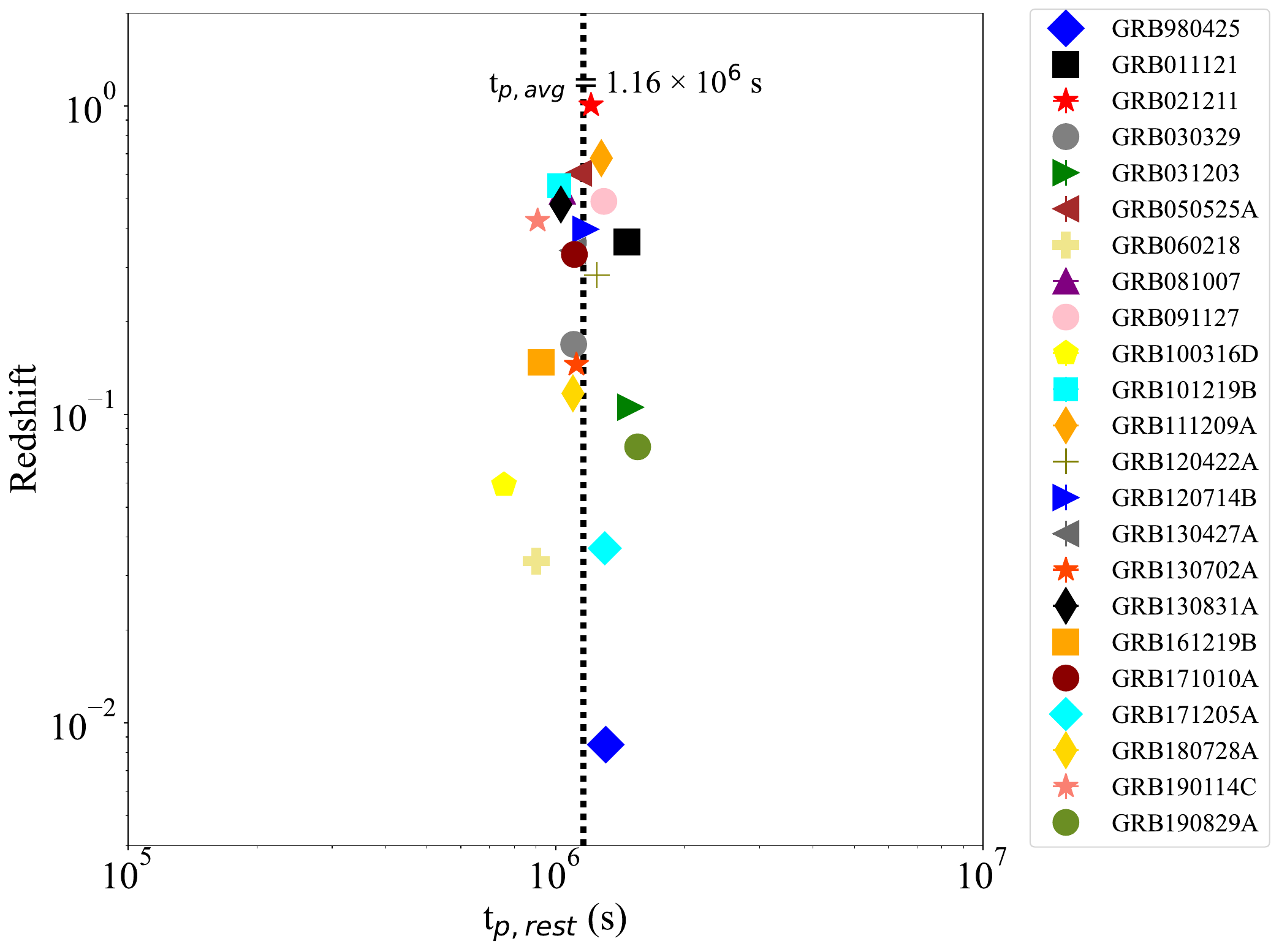}
    \includegraphics[width=.45\linewidth]{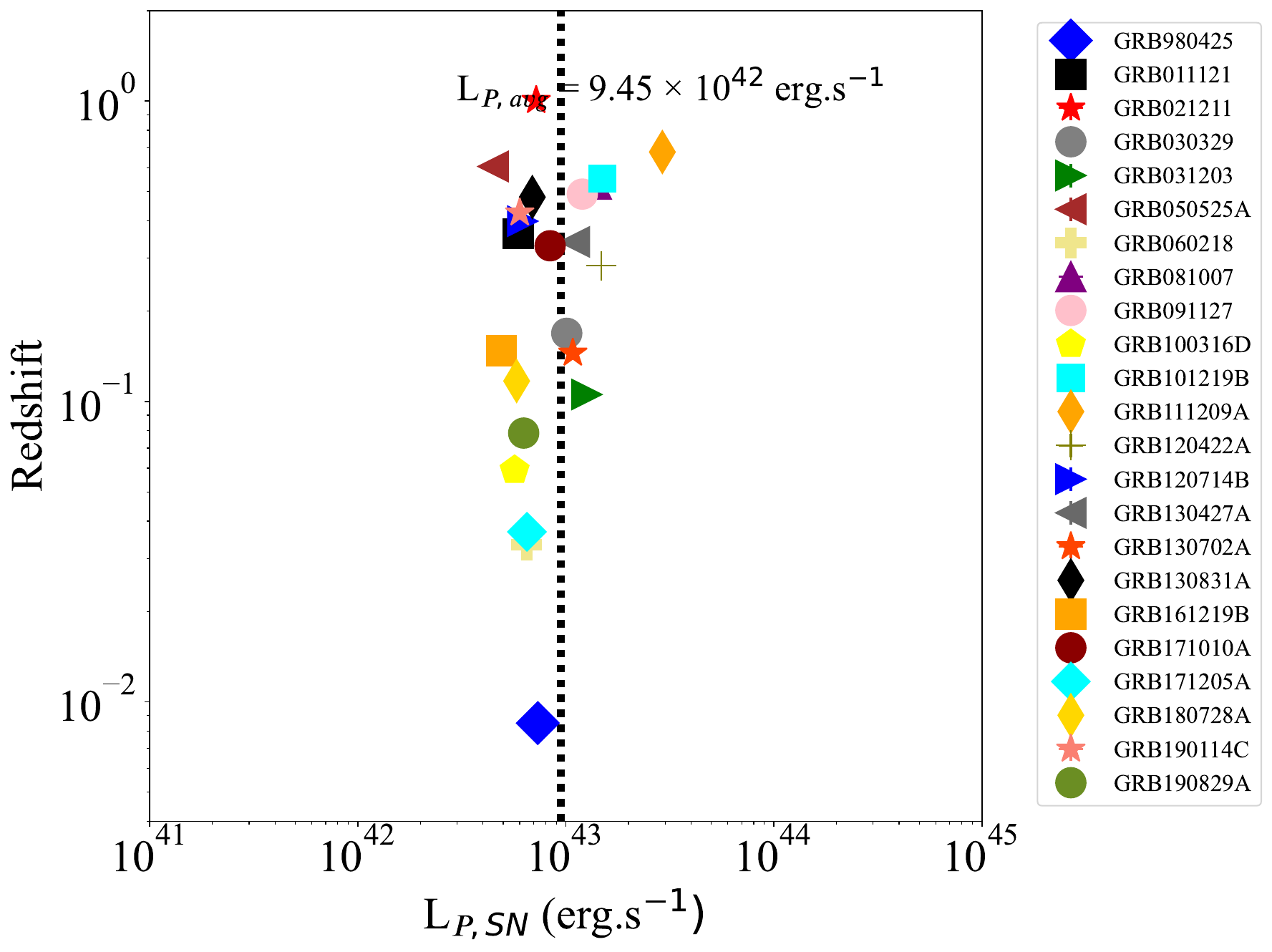}
    \caption{Left panel: GRB redshifts ($z$) versus the peak time of luminosity of the bolometric light curve of the associated SN ($t_{\rm P, SN}$). The plot shows the lack of correlation between these two quantities. Right panel: GRB redshifts ($z$) versus the values of peak luminosity of the bolometric light curve of the associated SN ($L_{\rm P, SN}$). The plot shows the spread in data points and the lack of correlation between these two quantities. Plots are taken from Ruffini et al. (2023).}
    \label{fig:Redshift-Tpeak-SN}
\end{figure}

Observations of supernovae (SNe) Ic occurring after the prompt emission of long gamma-ray bursts (GRBs) are addressed within the binary-driven hypernova (BdHN) model where GRBs originate from a binary composed of a $\sim10M_\odot$ carbon-oxygen (CO) star and a neutron star (NS). The CO core collapse gives the trigger, leading to a hypernova with a fast-spinning newborn NS ($\nu$NS) at its center. The evolution depends strongly on the binary period, $P_{\rm bin}$. For $P_{\rm bin}\sim5$min, BdHNe I occur with energies $10^{52}$--$10^{54}$erg. The accretion of SN ejecta onto the NS leads to its collapse, forming a black hole (BH) originating the MeV/GeV radiation. For $P_{\rm bin}\sim 10$min, BdHNe II occur with energies $10^{50}$--$10^{52}$erg and for $P_{\rm bin}\sim$hours, BdHN III occurs with energies below $10^{50}$erg. {In BdHNe II and III,} no BH is formed. The $1$--$1000$ms $\nu$NS originates, in all BdHNe, the X-ray-optical-radio afterglows by synchrotron emission. The hypernova follows an independent evolution, becoming an SN Ic, powered by nickel decay, observable after the GRB prompt emission. We report $24$ SNe Ic associated with BdHNe. Their optical peak luminosity and time of occurrence are similar and independent of the associated GRBs, see Figs. \ref{fig:Eiso-Tpeak-SN} and \ref{fig:Redshift-Tpeak-SN}. From previously identified $380$ BdHN I comprising redshifts up to $z=8.2$, we analyze four examples with their associated hypernovae. By multiwavelength extragalactic observations, we identify seven new Episodes, theoretically explained, fortunately not yet detected in galactic sources, opening new research areas. Refinement of population synthesis simulations is needed to map the progenitors of such short-lived binary systems inside our galaxy.

\begin{figure}
    \centering
    \includegraphics[width=\linewidth]{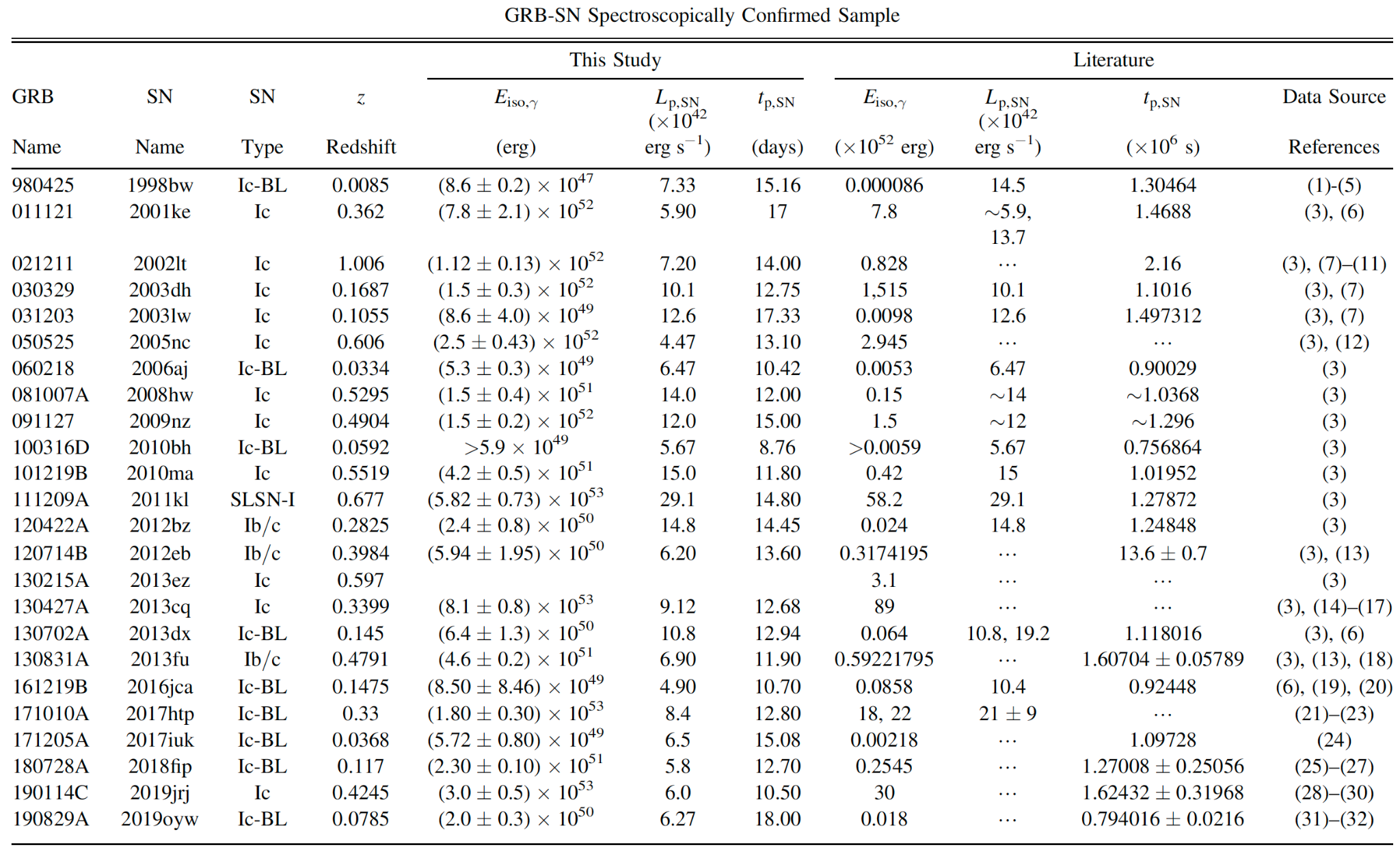}
    \caption{List of 24 spectroscopically confirmed GRB associated with SN. Taken from Ruffini et al. (2023).}
    \label{fig:24GRBSN}
\end{figure}





\section{GRB~220101A, GRB~221009A and GRB~160625B, and the seven episodes of BdHN I}\label{sec:part2-220101a-seven-episodes}

\begin{figure}
    \centering
    \includegraphics[width=\linewidth]{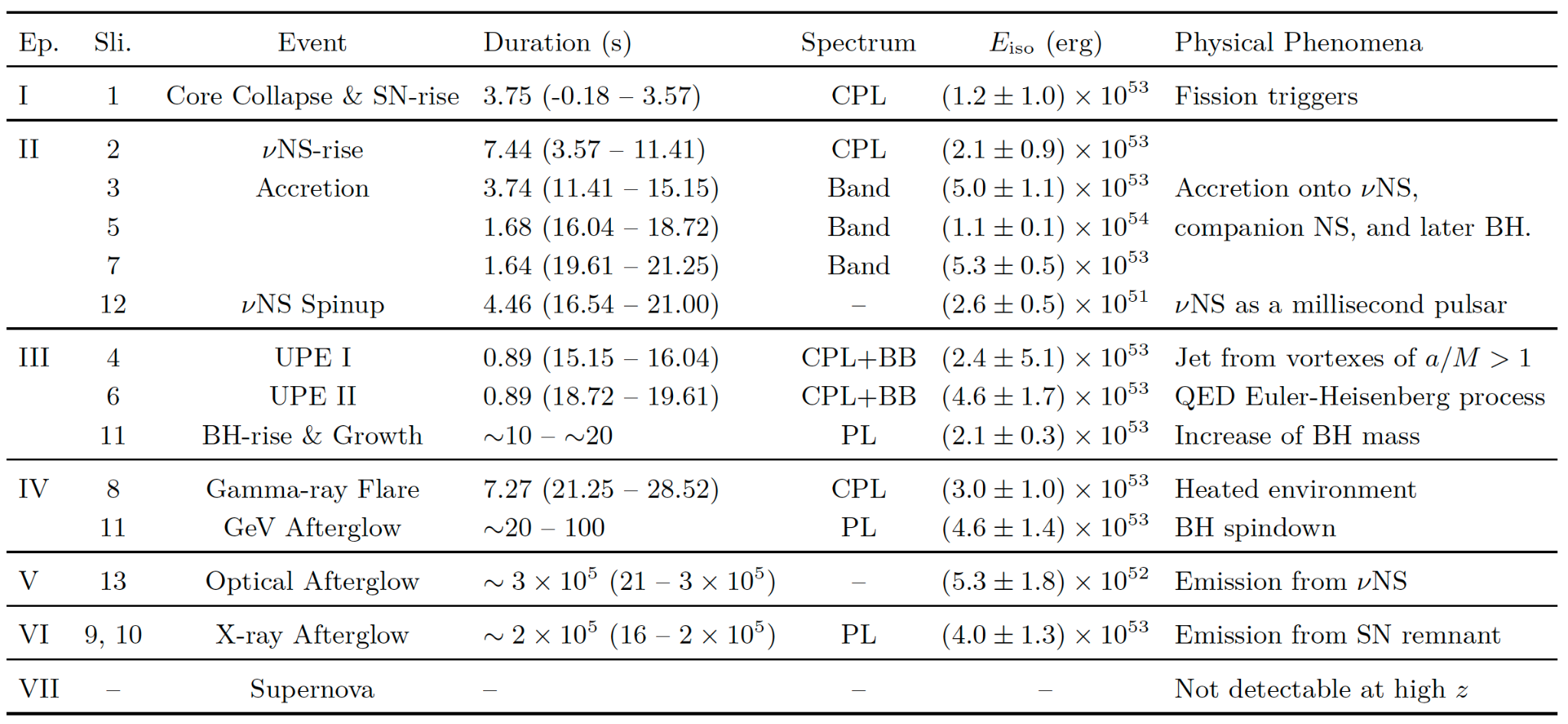}
    \caption{The episodes and afterglows of GRB 220101A. This table reports the episode name (Ep.), time slices (Sli.), duration (start–end in seconds), best-fit spectrum, isotropic energy in erg (Eiso), and underlying astrophysical processes for GRB 220101A. The redshift is $z=4.61$, and the total isotropic energy is $E_{iso}=4\times10^{54}$ erg Taken from Ruffini et al. (2023).}
    \label{fig:GRB220101A}
\end{figure}

\begin{figure}
    \centering
    \includegraphics[width=\linewidth]{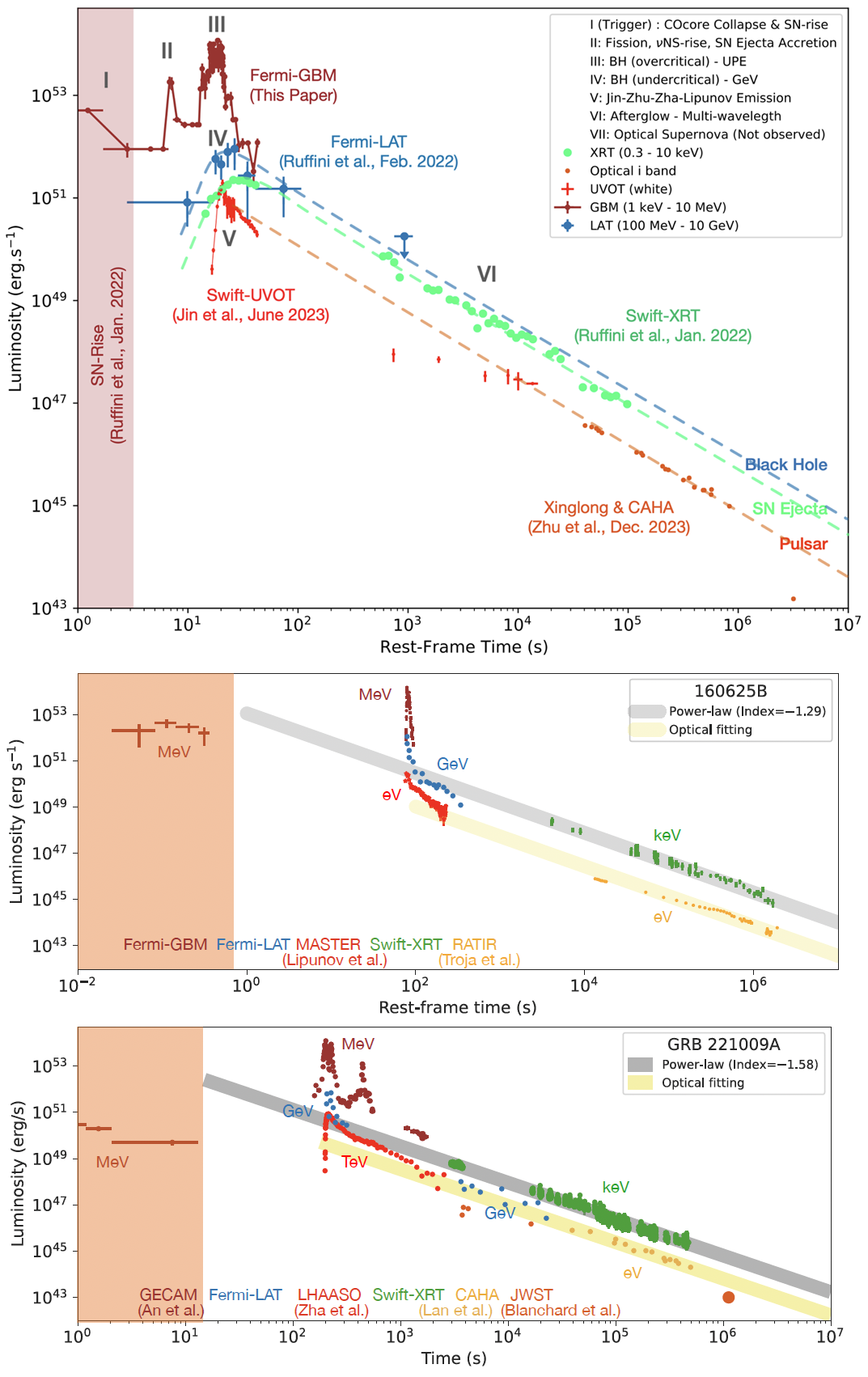}
    \caption{Multiwavelength light curve of GRBs 220101A, 160625B and 221009A with BdHN I episodes. Paper in preparation.}
    \label{fig:vhe221009A}
\end{figure}

On the ground of 24 spectroscopically confirmed GRB associated  with SN (see Fig. \ref{fig:24GRBSN}), we have identified seven different Episodes which we are 
further examining by inferring the astrophysical conditions which justify their occurrence. We acquired \cite{Ruffini24} the result that the early phase of the GRBs  is characterised by a CO core  and a binary NS companion.
The joint observations of the earliest phases of GRB220101A have manifested the highest ever initial luminosity of the Fermi GBM, and they have allowed to reach a deeper understanding of the UPE (1) and UPE (2) as vortices generated by the accretion of the SN ejecta onto the field of the companion  NS. 
Similarly due to the the high redshift of the source, the very early photons were recorded by the Swift XRT, originating the X-ray afterglow.
In addition to GRB220101A, see Fig. \ref{fig:GRB220101A}, we have also identified the seven Episodes characterizing  the GRB221009A and GRB160625B (see Fig. \ref{fig:vhe221009A} ).
The most remarkable new feature discovered in the GRB221009A has been the TeV radiation, coinciding with the birth of NS, and its spin-up by the pair-SN accretion.
Also remarkable are the slow velocities of accretion found both in GRB160625B and GRB221009A.
A new approach to this BDHN physics has been outlined in the paper of Ruffini, Mirtorabi, Fryer et al. (2025). A preliminary version is available.\cite{RR-Mirtorabi2025}
  
We are using these acquired knowledge in order to explore the physical evolution of the CO core and SN binary companion  which are naturally encountered in the origin of some new components e.g. the Supernova-Rise, which once introduced leads to even more extreme discoveries such as the possible existence as the triggering in the BDHN1 model, of a sequence of a pair SN, followed by a Core collapse SN.

In addition the beaming present in the GeV component of the of a BDHNe can lead to a special understanding on the relative manifestation in the structure of the remnant to a particular ratio of the radio, X-ray  and GeV TeV emission of the  remnant.

\section{GRB~221009A and Very High-Energy photons}
\label{sec:part2-221009a-vhe-photons}

In addition, very important results have been obtained from GRB 221009A, strikingly confirming the luminosity and time sequence of the TeV data. This source has also provided us with the opportunity to observe SN, which triggered the event, as described above, see also Fig. \ref{}.

\begin{figure}
    \centering
    \includegraphics[width=\linewidth]{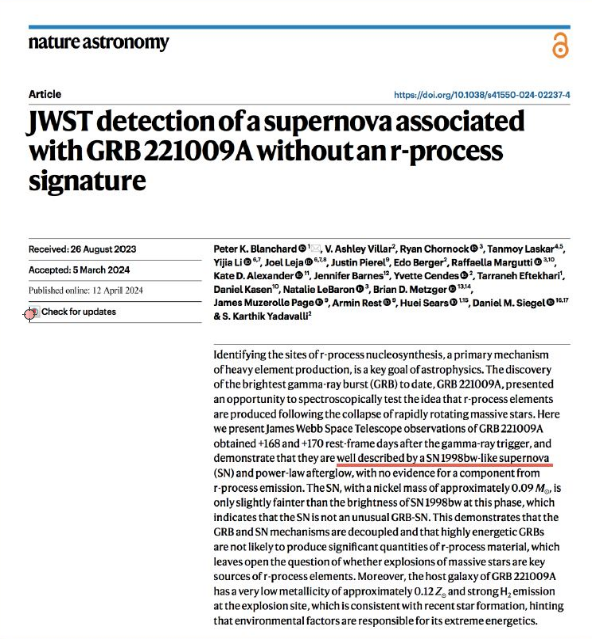}
    \caption{Optical identification by the JWST of the afterglow emission in GRB 221009A.}
    \label{fig:JWST221009A}
\end{figure}











Before closing I would like to return on the initial working hypothesis that the data of GRB190114C could be considered, as soon as extrapolated to $10^3$ years after the trigger, a source analogous to the Crab Nebula remnant and they could have been used in reaching a deeper understanding of the recorded appearance arena of the 1054 observations in the Rocky Mountains as well as a catastrophic appearance of the consequences of the supernova explosion in the Mediterranean region and in Egypt, traditionally considered as a manifestation of a plague and reinterpreted as a natural "Hiroshima effect", see e.g. \cite{RR-Sigismondi2024}.

\begin{figure}
    \centering
    \includegraphics[width=0.8\linewidth]{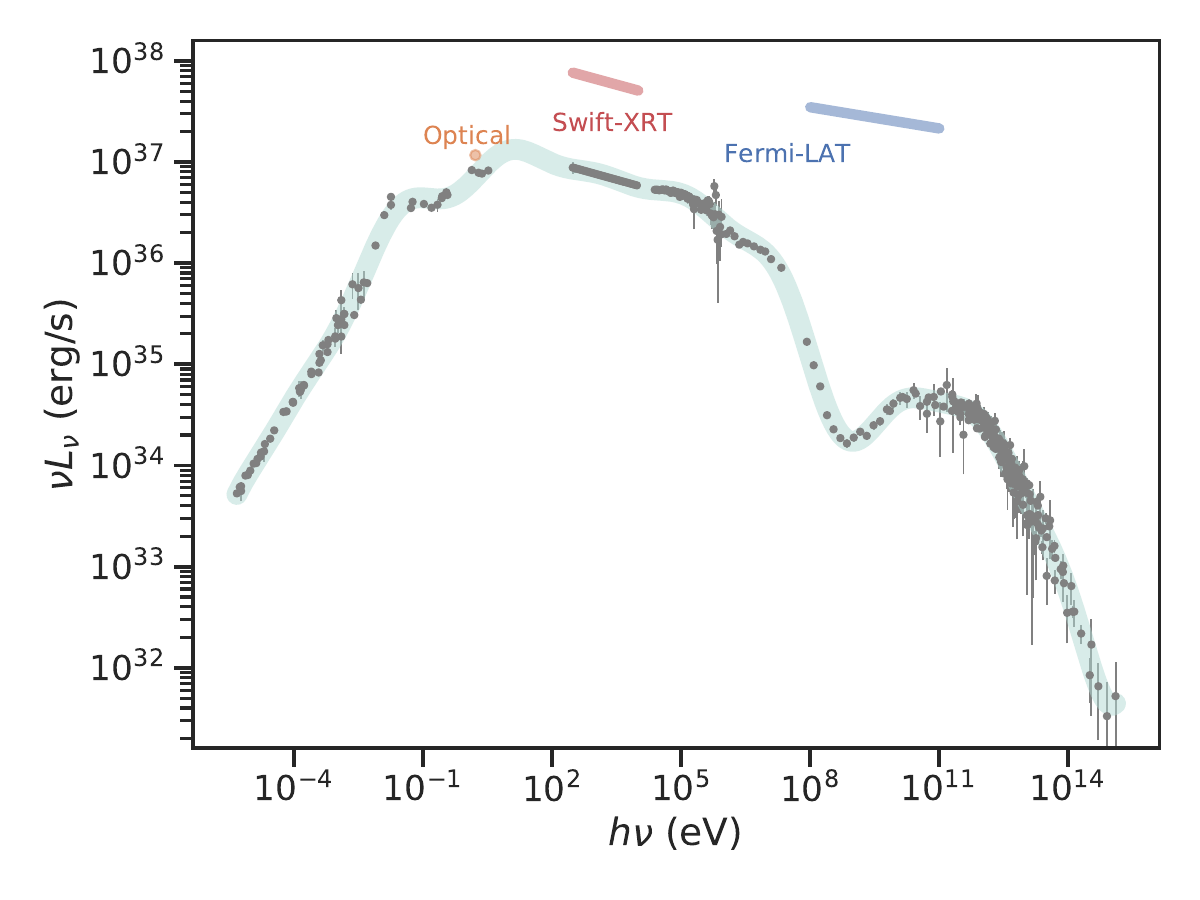}
    \caption{Multiwavelength spectrum of the Crab nebula with afterglow imprint from GRB 220101A,\ref{fig:220101A-Crab-Spectrum} (see also Fig. \ref{fig:crab_190114c_imprint} with the same operation made with GRB 190114C). The agreement between optical (yellow) and X-ray (red Swift XRT) afterglows extrapolations after 971 years is fairly good, but the GeV extrapolated radiation (blue Fermi-LAT) is more than two orders of magnitude than the presently observed one. Also this case is possibly explained by a beaming effect of the BdHN GeV-source for the Crab, which is not aligned with our line of sight.}
    \label{fig:220101A-Crab-Spectrum}
\end{figure}
But what is truly remarkable is that figure \ref{fig:220101A-Crab-Spectrum} representing the effects of GRB190114C reinterpretation is perfectly consistent with the corresponding ones of GRB 220101 (courtesy of Prof. Wang Yu). What is really tantalizing is to conclude that all long GRBs have a common development after $10^3$ years after the trigger in a Remnant similar to the Crab Nebula and the nature of that spectra can be explained in term of three afterglows: one in the optical, leading to a 30-40 ms pulsar, one in the GeV originating from the BH formation, and finally  an X-ray afterglow originating from the synchrotron radiation emission identified in the Crab remnant identified in the classic book of Shklovsky \cite{1969supe.book.....S} as well as by  the James Webb Space Telescope \cite{2024ApJ...968L..18T}.

\section{Conclusions}\label{sec:part2-conclusions}

The observations of GRBs in extragalactic sources are indeed essential in identifying the seven independent Episodes of a GRB, extending our current knowledge of the fundamental laws of physics. This allows, among others, to explain the origin of the physical component of our Universe, essential to the life on our planet, e.g., the  existence of Nickel-Cobalt abundance. This opens the way to the understanding of the possible roles of GRBs, including, e.g., the origin of DNA.

We have shown that the multi-messenger observations of the Crab SN remnant, both the historical and current ones, can be explained in terms of a GRB similar to GRB 190114C with $10^{54}$~erg. This has allowed us to reconstruct the first appearance of the Crab SN on July~3,~1054, with observations from the Rocky Mountains and the ``Hiroshima-like'' events over Egypt at Noon. This proves that the occurrence of a GRB inside our own galaxy is possible without destroying life on the Earth, at $\approx2$~Kpc distance.

A new process is just starting: to use all data, especially the new physical laws characterizing each of the seven episodes to infer the nature of SNe, triggering the GRBs. The goal is not just to fit the SN energy by quantitative explanation, but explaining out of the first principles the extraordinary energy of SN rise, first tentatively introduced in Ruffini et al. (2021).\cite{RR-MNRAS_2021} The clear evidence for the first observation of the pair SN followed a few seconds later by a normal core-collapse SN appears to be the main difference in this new just born scenario, see Ruffini, et al. (2025, submitted).\cite{RR-Mirtorabi2025}

\section{Latest developments on X-fermion}

Following the publication of ``Introducing the black hole'' \cite{1971PhT....24a..30R} Charles Townes and Reinhard Genzel launched a proposal to look for a Kerr BH in our Galactic Center: Sgr A*. The first results appeared already in 1987 by Genzel and Townes addressing the dynamics and mass distribution in the center of our Galaxy \cite{1987ARA&A..25..377G}. Nobody may have expected the tremendous progress in new technology and astrophysical knowledge such a proposal will generate in the following 38 years. The nucleus of our Galaxy was further reviewed by Genzel, Hollenbach and Townes in 1994 \cite{1994RPPh...57..417G}, with the evidence that the motion of a system of stars orbiting Sgr A* determines the mass $M_{SgrA^*}$ of the central object by Genzel, Eckart, Ott and Eisenhauer in 1997 \cite{1997MNRAS.291..219G}, the evidence for flaring activity at the Galactic Center by Genzel, Sch\"odel, Ott, et al. \cite{2003Natur.425..934G} soon followed, as well as the paper by Genzel, Eisenhauer and Gillessen \cite{2010RvMP...82.3121G} focusing on the faint young “S-star cluster” around Sgr A*.
One of the most important discoveries in the last decade obtained thanks to the high quality data from the GRAVITY interferometer at VLT \cite{2018A&A...615L..15G,2020A&A...636L...5G} and Keck Observatory \cite{2019Sci...365..664D} have been the measurement of the gravitational redshift and perihelion precession of the S2 star in the core of Sgr A* . These results firmly establish the mass $M_{SgrA^*}=4.3\times 10^6 \,M_{\odot}$ \cite{2024A&ARv..32....3G} in the Galactic center.

\begin{figure}[ptb]
\centering
\includegraphics[width=\columnwidth]{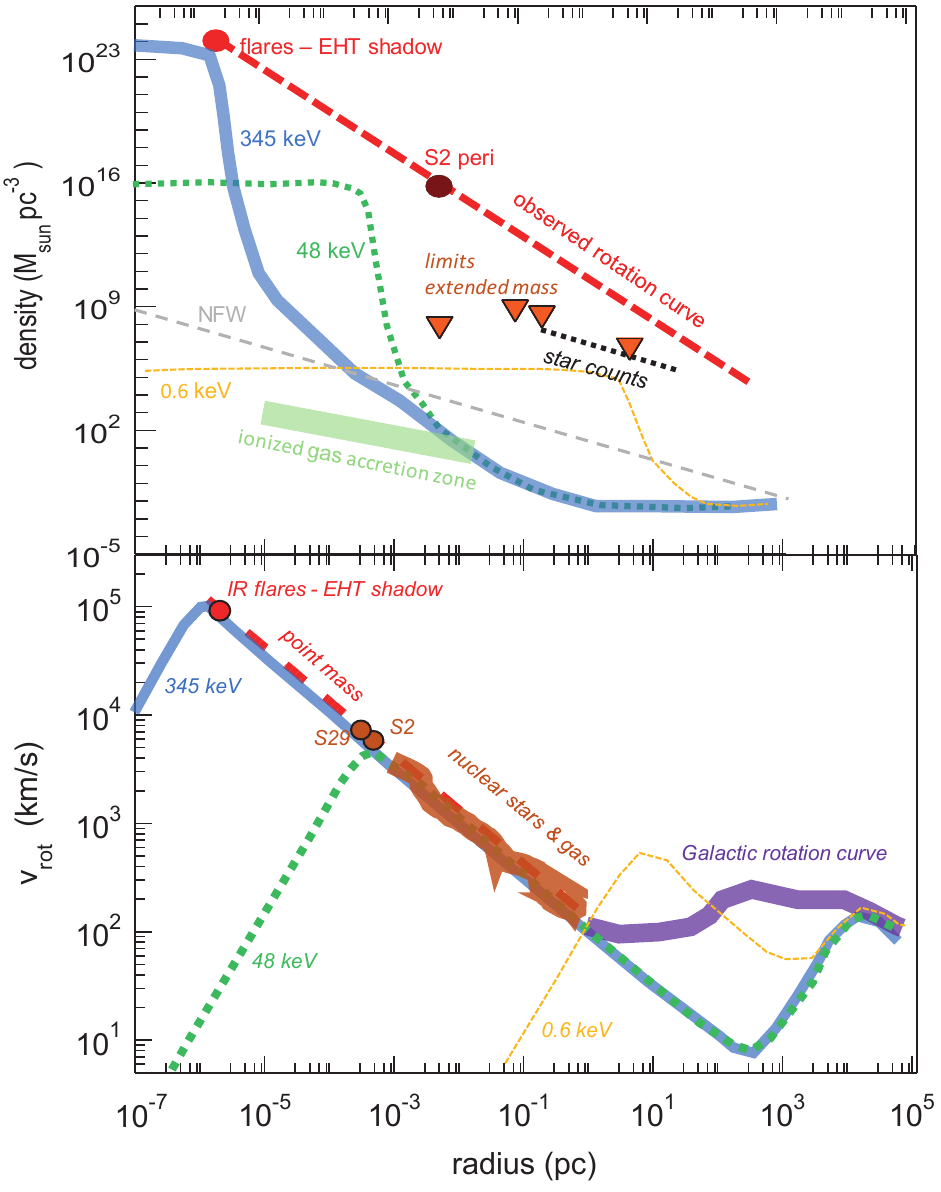}
\caption{Constraints on the contribution of various mass components of baryonic and dark fermionic matter to the mass density (left) and rotation velocity (right) at different radii $R$. Reproduced by courtesy from Genzel, Eisenhauer and Gillessen (2024)
. We are grateful to Prof. Reinhard Genzel for giving us permission to reproduce his figure.}%
\label{Fig7-limits-fermion-ball}%
\end{figure}

In the meantime Ruffini and collaborators, based on earlier studies of self-gravitating equilibrium configurations \cite{1989A&A...221....4M,1990A&A...227..415M,1990A&A...235....1G}, and the first attempts to fit the rotation curves of galaxies \cite{1980ApJ...238..471R,1996MNRAS.281...27P,1986A&A...157..293M}, have developed the Ruffini-Arg{\"u}elles-Rueda (RAR) model treating the DM in galaxies as a system of self-gravitating fermions at finite temperatures. They distinguish two components: a degenerate fermionic core with typical dimensions of $R_{core}\sim 10^{-3}\mbox{--}10^{-6}$ pc and extended halo with size $R_{halo}\geq 10$ kpc described by the Boltzmann statistics, see \cite{2015MNRAS.451..622R,2018PDU....21...82A} and Fig. \ref{Fig7-limits-fermion-ball} reproduced from review \cite{2024A&ARv..32....3G}. This model determines the overall distribution of DM in galaxies by solving the equations of equilibrium at selected temperatures and chemical potential within General Relativity. It has been applied to our Galaxy \cite{2018PDU....21...82A} and other galaxies \cite{2019PDU....24..278A,2023ApJ...945....1K}, for a review see \cite{2023Univ....9..197A}. One of the general conclusions from the analysis of rotation curves of galaxies and especially of our Galaxy is that both core and halo are essential and are at the basis of galactic structures. The mass of the core depends on particle mass, while the mass of the halo remains unconstrained, unless a cut-off in its momentum distribution is implemented \cite{1989A&A...221....4M,1990A&A...235....1G,2018PDU....21...82A}.

The interpretation of the data from \cite{2020A&A...636L...5G} within the RAR model \cite{2020A&A...641A..34B,2021MNRAS.505L..64B} allowed Becerra-Vergara et al. to explain the relativistic effects in the motion of the S cluster stars and the G2 object without invoking the presence of a BH: they originate in the degenerate core of selfgravitating system of DM fermions. A lower limit on the fermion mass was derived: $m_X>56$ KeV \cite{2020A&A...641A..34B,2022MNRAS.511L..35A}. What has been very remarkable is that the same value of the fermion mass can be used to simultaneously fit the data from our Galactic core observed by GRAVITY and Event Horizon Telescope (EHT), as well as the halo data obtained from Gaia, see Fig. \ref{Fig7-limits-fermion-ball} and \cite{2025arXiv250310870K,2025arXiv251019087C}. While in our work we use physically motivated choice based on the Fermi-Dirac statistics, several alternative approaches to the core-halo (or central BH versus DM halo) problem exist. Usually a given density profile and equation of state are assumed in order to determine the properties of the configuration. For instance, for anisotropic fluid de-Sitter-like equation of state the central object was studied assuming Einasto profile \cite{2021EPJC...81..777B} and Zaho profile \cite{2022EPJC...82..759B}. In \cite{2019MNRAS.484.3325B} the exponential sphere profile was adopted for the core and the NFW one for the halo.

A major difference emerges from the above treatment by comparing and contrasting the baryonic and the DM distribution: the DM distribution both in the core and in the halo has only gravitational interactions and zero net angular momentum. It is dominated by the sole Fermi quantum statistics generalized to the fully general relativistic regime. All the other interactions of the standard model, including the Fermi weak interactions are present only in the baryonic matter component. Among these the galactic angular momentum which has a special role in the explanation of the nature of AGN.

We recall that the critical mass for the gravitational collapse to a BH is only a function of the mass $m_X$ of the fermion
\cite{1939PhRv...55..374O,2023MNRAS.523.2209A,2024ApJ...961L..10A}
\begin{equation}
M_{cr}\simeq0.384\frac{M_{P}^{3}}{m_X^{2}}=6.95\times10^{6}\left(\frac{300\,\mathrm{keV}}{m_X}\right)  ^{2}\,M_{\odot},
\label{Mcr}%
\end{equation}
where $m_{P}=\sqrt{\hbar c/G}$ is Planck's mass. For $m_X=56$ KeV the critical mass (\ref{Mcr}) is $M_{cr}=2.0\times 10^8 \,M_{\odot}$.

From the above theoretical considerations and observational constraints we infer the absolute lower limit to a cosmological DM SMBH in \emph{any} galaxy:
\begin{equation}
M_{\text{SMBH}}>M_{SgrA^*}=4.3\times 10^6 \,M_{\odot}.
\label{Mmin}
\end{equation}
as well as $m_X<381$ KeV.

No matter the value of $m_X$, our Galactic Center will necessarily evolve, by accretion process, to form a BH. As we will see in the conclusions, this determines a ``\emph{minimum time interval of stability}'', which is a function of the actual mass of the core, of the mass accretion rate $\dot{M}$ and the mass of X-fermion $m_{X}$.

But let us turn to the more general quest of the presence of DM versus baryonic matter in the cosmological large scale structure.
Slowly but conclusively evidence was gained that: a) DM distribution at galactic scales and on cosmological scales is different from the distribution of baryons. It includes the mass-to-light ratio measurements in individual galaxies $M/L\sim 10\,M_{\odot}/L_{\odot}$ and in clusters of galaxies $M/L\sim 300\,M_{\odot}/L_{\odot}$ \cite{2008gcpa.book.....W}, the measurement of their luminosity functions \cite{2001AJ....121.2358B,2009MNRAS.399.1106M}, mapping the mass distribution from the data of weak and strong gravitational lensing \cite{2010RPPh...73h6901M,2023arXiv230611781V} with dedicated space observatories such as Euclid mission \cite{2024arXiv240513491E} and the determination of cosmological parameters from the CMB radiation fluctuations by such missions as Planck \cite{2020A&A...641A...6P}, which gives the density parameter for baryons $\Omega_b h^2 = 0.0224$ and for the DM $\Omega_{DM} h^2 = 0.120$, where $h_{0}=0.674$\ is the dimensionless Hubble parameter; b) possibly the greatest revolution in galactic structures is the discovery by the JWST of the LRDs \cite{2024ApJ...963..129M,2024arXiv241114383R,2024arXiv241102729Z}, harboring SMBHs in the range $10^{6}\mbox{--}10^{9}\,M_{\odot}$ with sizes $<100$ pc all the way to redshift $z\sim 12$; c) the presence of SMBHs with masses $M>10^{9}\,M_{\odot}$ in Quasars at $z\sim2$ and at high redshifts $z>6$ \cite{2021ApJ...923..262Y,2023ARA&A..61..373F}.

With reference to LRDs we propose that these systems represent a successive era to the AGNs and define this era as the Quiescent Galactic Nuclei (QGN) era, when the energy source of AGN has been exhausted and no further emission in X-ray or high energy occurs. These QGNs are dominated by the accretion of baryonic matter onto the quantum DM degenerate core fulfilling the mass constraint determined in (\ref{Mmin}).
In a separate publication \cite{RuffiniDellaValleWang2025,Ruffini2025a} the data of observed Quasars J0313-1806 at $z=7.642$, J010013.02+280225.8 $z=6.30$ and UHZ-1 at $z\sim 10$ \cite{2024ApJ...961L..10A} are extrapolated to an earliest phases triggered by a DM distribution given by (\ref{Mmin}) in the case of LRDs differentiating between the case of AGN dominated by Eddington luminosities evolving in AGN, and the case dominated by Bondi accretion phase characterizing a QGN.

Our main interest is to determine the origin of the galactic structures with a) a nonrotating degenerate DM core leading to, or already harboring a BH, b) a nonrotating extended halo of X-fermions following the Boltzmann statistics as well as c) a rotating baryonic matter, which by accretion in the above DM core can explain the nature of AGNs with distinctive X-ray and high energy emission \cite{2013ARA&A..51..511K,2024ApJ...961L..10A}.

Possibly related to the nature of the X-fermion is the renewed interest in the electron-positron annihilation line \cite{2005MNRAS.357.1377C,2005A&A...441..513K} observed from our Galactic Center and the anomalous ionization rate observed in the Central Molecular Zone, attributed to DM annihilation into electron-positron pairs \cite{PhysRevLett.134.101001} for particle mass in the range $1\, \text{MeV} < m< 1\, \text{GeV}$.

Decades of progress culminating in the optical interferometry developed in the Keck Observatory and VLT GRAVITY facility led to the determination of the mass of the compact object in Sgr A* in $4.3\times 10^6 \,M_{\odot}$ and firm limits on its angular momentum. The most straightforward interpretation of these observations is a Schwarzschild black hole (BH). Considering that the S cluster of stars orbiting the Galactic Center does not probe distances closer than 1400 its Schwarzschild radius, several alternatives were proposed. Among them the self-gravitating system of Dark Matter (DM) fermions (here referred to as \emph{X-fermion}) having a quantum degenerate core and extended to non-degenerate isothermal halo. Application of this model to our Galaxy results in a bound on X-fermion mass $m_X>56$ KeV. Upon accreting baryonic matter the degenerate fermionic core may collapse into a BH providing seeds to Active Galactic Nuclei (AGNs) and SuperMassive Black Holes (SMBHs).
We address the implications of DM X-fermions with mass $m_X\sim 300$ keV endowed with a large and negative chemical potential here evaluated on structure formation and its evolution with redshift. The Jeans mass of X-fermions peaks at $10^{10}\,M_{\odot}$ at redshift $z\sim 10^9$, which sets the scale of galactic structures. The degenerate cores forming as early as $z\sim 14$ are consistent with recent observations of the James Web Space Telescope (JWST) disclosing the Little Red Dots (LRDs) at cosmological redshifts  $4<z<12$ harboring SMBHs in the range $10^6\mbox{--}10^9\,M_{\odot}$.
Assuming that Sgr A* is not yet a BH, the lower limit on SMBH mass of $4.3\times 10^6 \,M_{\odot}$ and the corresponding upper limit on the X-fermion mass of $m_X<381$ KeV are established, which acquires relevance given ongoing JWST observations of the Galactic Center and the planning of new accelerators for the scrutiny of DM in CERN.

\section{On gravitational waves and the period following 1985 in Sapienza. The verification of absence of validity of Weber's results obtained in 17 years of cryogenic detectors AURIGA, NAUTILUS, EXPLORER and ALLEGRO}

After his graduation in "La Sapienza" (March 1967) and a postdoctoral position in Pasqual Jordan’s group (April-August 1967), Ruffini departed for Princeton (September 1971-1972), working with John Archibald Wheeler. From 1973-1976 he was an assistant Professor at Princeton University and Stanford University, a Member of the Institute for Advanced Study IAS and was nominated an Alfred P. Sloan Foundation fellow. He collaborated with Wheeler for “Relativistic cosmology from Space platform” (ESO SP52), later reproduced in the first  10 chapters of the book “Black Holes, Gravitational Waves and Cosmology” by Rees-Ruffini-Wheeler (1974). The excerpts from ESO SP52 motivated the paper Ruffini-Wheeler “Introducing the Black Hole” (Physics Today, January 1, 1971). Soon after followed the formulation of the black hole mass energy formula (1) September 17, 1970: Christodoulou, Phys. Rev. Lett., 25 (1970) 1596; followed by 2) March 1, 1971: Christodoulou, Ruffini, Phys. Rev. D, 4 (1971) 3552; followed by 3) March 11, 1971: Hawking, Phys. Rev. Lett., 26 (1971) 1344). The final mass energy formula reads:
\begin{gather}
M^2=\left(M_{irr}+\frac{e^2}{4M_{irr}}\right)^2+\frac{L^2}{4M_{irr}^2}.
\end{gather}

In 1972, the theoretical work on the absolute upper limit to the neutron star mass (see e.g. Rhoades, E.; Ruffini R., Maximum Mass of a Neutron Star, Phys. Rev. Lett. 32, 324 (1974) led to the identification of Cygnus X-1 as the first black hole in our Galaxy, presented at the 6th Texas Symposium, and soon after recognized with the Cressy Morrison Award (Transactions of the New York Academy of Sciences, March 1973).

At Stanford, during his collaboration on Gravity Probe B (PI F. Everitt), and following the Hulse and Taylor discovery of the first binary pulsar, Damour and Ruffini (1974) used PSR B1913+16 to test spin orbit precession in general relativity, predicting its disappearance in 2025, now under observational test (Kramer, ApJ 509, 856–860). Their work on higher order relativistic effects provided quantitative confirmation of gravitational wave emission (Taylor, Fowler, and McCulloch, 1979).

At the close of his United States period, Ruffini co-authored with Riccardo Giacconi Physics and Astrophysics of Neutron Stars and Black Holes (North Holland, 1978), a volume involving seven Nobel laureates. Internationally, he published with H. Sato in Japanese (Chuo Koron Sha, Tokyo, 1976), with Ohanian in Korean (W. W. Norton and Shin Won Agency, 2001), and founded research groups in Australia. He was the first Western astrophysicist to visit Maoist China in 1976, beginning fifty years of collaborations and joint meetings. In 1978 he returned to Italy, first in Catania, then at Sapienza as chair in theoretical physics, a position previously held by Enrico Fermi.

In Italy, Ruffini continued after 1985 to collaborate with the United States (Stanford, Space Telescope Science Institute, University of Washington), with the Vatican Observatory, and with ICTP Trieste through ICRA at Sapienza, later extending to Armenia and the Holy See with the establishment of ICRANet in 2005. He promoted continuous international data sharing, starting with the Wide Field X ray Camera for BeppoSAX. With collaborators he advanced from binary X ray sources (Gursky and Ruffini, Springer, 1975) to the establishment of the seven episode GRB supernova association in GRB 220101 (2025).

Before going to the US, Ruffini had promised Amaldi to report on  Weber’s gravitational wave detections, published in five PRL (1966 and 1973) without any astrophysical counterparts. The collaboration through ICRA (since 1985) and ICRANet (since 2005), with Amaldi, Fairbank, and their research groups led to advanced cryogenic detectors at 3 mK (AURIGA, NAUTILUS, EXPLORER, ALLEGRO) with higher sensitivity. No Weber events were found during thirty years of observations as reported (IGEC 2, P.R.D. 76, 102001, 2007, 2013), see also 10 papers, mainly in Phys. Rev. D, by Pia Astone et al. from 1996 up to 2013.

In conclusions, the Weber’s events were due to an incorrect statistical treatment of thermal noise:

\begin{itemize}
    \item Weber’s detections were incorrect;
    \item a reassessment of works connected to Weber’s observations is required;
    \item The considerations on the surface area theorem by Hawking have also to be reassessed and, in particular, the difference between this Hawking theorem and the role of the Mirr. This difference, in the meantime, has been made explicit in 1) R. Ruffini et al., Single versus the Repetitive Penrose Process in a Kerr Black Hole, Phys. Rev. Letters 134, 081403 (2025); 2) Ruffini et al., Role of the irreducible mass in repetitive Penrose energy extraction processes in a Kerr black hole, Phys Rev Research 7, 013203 (2025) and 3) Ruffini R., Zhang S., in preparation. See summary in “The Black Hole Mass-Energy Formulae”.
\end{itemize}

\begin{figure}
    \centering
    \includegraphics[width=0.5\linewidth]{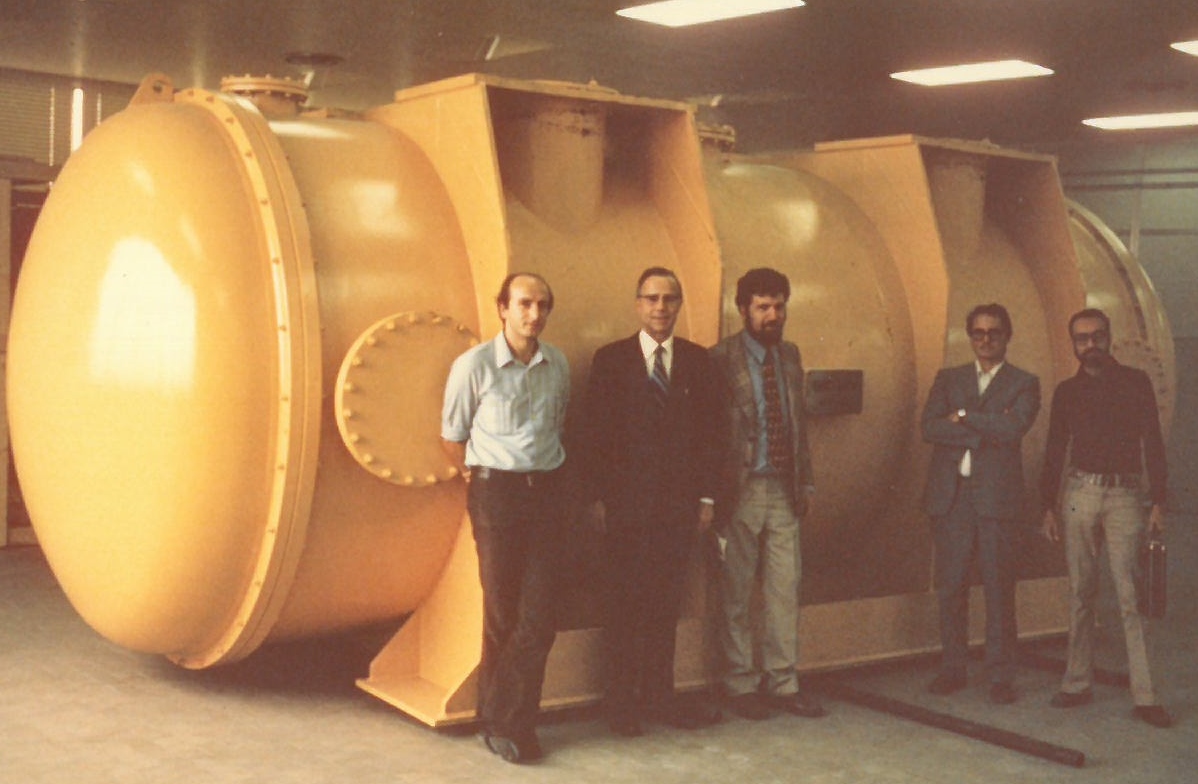}
    \caption{Gravitational Wave detector NAUTILUS. At the center Remo Ruffini, on his right William Fairbank, on his left Guido Pizzella.}
    \label{fig:Nautilus1}
\end{figure}
\begin{figure}
    \centering
    \includegraphics[width=0.5\linewidth]{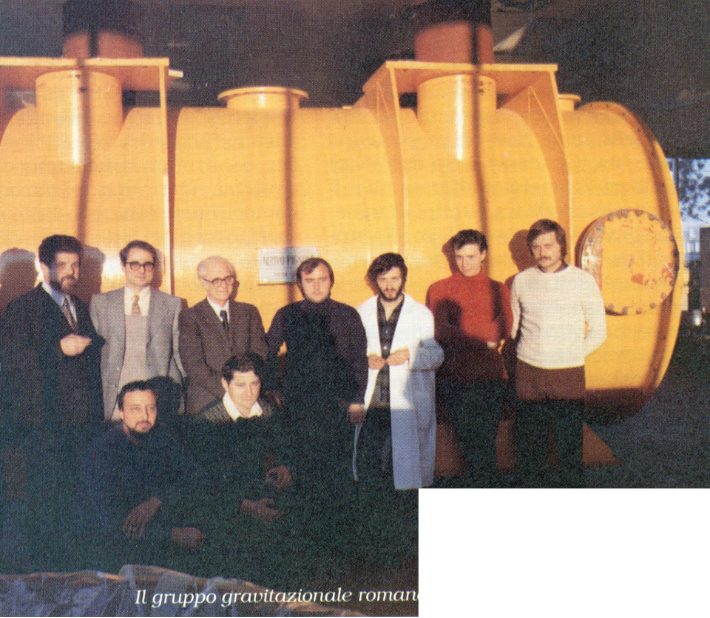}
    \caption{Gravitational Wave detector NAUTILUS. From the left to the right: Remo Ruffini, Guido Pizzella, Edoardo Amaldi.}
    \label{fig:Nautilus2}
\end{figure}
\begin{figure}
    \centering
    \includegraphics[width=0.5\linewidth]{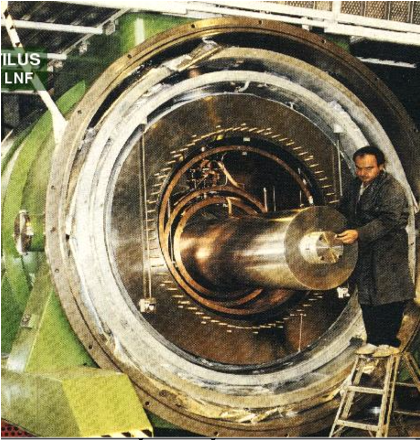}
    \caption{The Nautilus cryogenic detector.}
    \label{fig:Nautilus3}
\end{figure}

No Weber’s Gravitational Wave Signals at $10^{-3}$ K from 1996 to 2013 in the Criogenic Detectors  AURIGA, EXPLORER, NAUTILUS, see Figs. \ref{fig:Nautilus1},\ref{fig:Nautilus2},\ref{fig:Nautilus3}.

\begin{enumerate}

    \item P. Astone, M. Bassan, P. Bonifazi , P. Carelli, E. Coccia, C. Cosmelli, V. Fafone, S. Frasca, A. Marini, G. Mazzitelli, I. Modena,  G. Modestino, A. Moleti, G.V. Pallottino, M.A. Papa, G. Pizzella, P. Rapagnani, F. Ricci, F. Ronga, R. Terenzi, M. Visco, L. Votano, Upper limit for a gravitational-wave stochastic background with the EXPLORER and NAUTILUS resonant detectors (Physics Letters B 385, 1996)
    \item P. Astone, M. Bassan, P. Bonifazi, P. Carelli, E. Coccia, C. Cosmelli, S. D’Antonio, V. Fafone, S. Frasca, Y. Minenkov, I. Modena, G. Modestino, A. Moleti, G. V. Pallottino, M. A. Papa, G. Pizzella, L. Quintieri, F. Ronga, R. Terenzi, M. Visco, Search for periodic gravitational wave sources with the Explorer detector (Physical Review D, volume 65, 022001, 2001)
    \item P. Astone, M. Bassan, P. Bonifazi, P. Carelli, G. Castellano, E. Coccia, C. Cosmelli, G. D’Agostini, S. D’Antonio, V. Fafone, G. Federici, F. Frontera, C. Guidorzi, A. Marini, Y. Minenkov,  I. Modena, G. Modestino,  A. Moleti, E. Montanari, G. V. Pallottino, G. Pizzella, L. Quintieri, A. Rocchi, F. Ronga, R. Terenzi, G. Torrioli, M. Visco, Search for correlation between GRB’s detected by BeppoSAX and gravitational wave detectors EXPLORER and NAUTILUS (Physical Review D 66, 102002, 2002)
    \item P. Astone,  D. Babusci, L. Baggio, M. Bassan, D. G. Blair, M. Bonaldi, P. Bonifazi, D. Busby, P. Carelli, M. Cerdonio,E. Coccia, L. Conti, C. Cosmelli, S. D’Antonio,V. Fafone, P. Falferi,P. Fortini, S. Frasca,G. Giordano, W. O. Hamilton,I. S. Heng, E. N. Ivanov,W. W. Johnson, A. Marini, E. Mauceli, M. P. McHugh, R. Mezzena, Y. Minenkov, I. Modena,G. Modestino, A. Moleti, A. Ortolan,G. V. Pallottino, G. Pizzella,G. A. Prodi, L. Quintieri, A. Rocchi, E. Rocco, F. Ronga, F. Salemi, G. Santostasi,L. Taffarello, R. Terenzi, M. E. Tobar, G. Torrioli,G. Vedovato, A. Vinante, M. Visco, S. Vitale, J. P. Zendri, Methods and results of the IGEC search for burst gravitational waves in the years 1997–2000, (Physical Review D 68, 022001, 2003).
    \item P. Astone, D. Babusci, M. Bassan, P. Carelli, E. Coccia, C. Cosmelli,  S. D’Antonio,V. Fafone,F. Frontera,G. Giordano, C. Guidorzi,  A. Marini,  Y. Minenkov,  I. Modena,  G. Modestino, A. Moleti, E. Montanari, G.V. Pallottino, G. Pizzella, L. Quintieri, A. Rocchi, F. Ronga, L. Sperandio, R. Terenzi, G. Torrioli, M. Visco, Cumulative analysis of the association between the data of the gravitational wave detectors NAUTILUS and EXPLORER and the gamma ray bursts detected by BATSE and BeppoSAX, (Physical Review D 71, 042001,2005)
    \item P. Astone, D. Babusci, L. Baggio, M. Bassan, M. Bignotto,M. Bonaldi, M. Camarda, P. Carelli, G. Cavallari, M. Cerdonio, A. Chincarini, E. Coccia, L. Conti, S. D’Antonio,  M. De Rosa, M. di Paolo Emilio, M. Drago, F. Dubath,V. Fafone, P. Falferi, S. Foffa, P. Fortini, S. Frasca, G. Gemme, G. Giordano, G. Giusfredi, W. O. Hamilton,  J. Hanson, M. Inguscio, W.W. Johnson, N. Liguori, S. Longo, M. Maggiore, F. Marin, A. Marini, M. P. McHugh, R. Mezzena, P. Miller,Y. Minenkov, A. Mion, G. Modestino, A. Moleti, D. Nettles, A. Ortolan, G.V. Pallottino, R. Parodi,G. Piano Mortari, S. Poggi, G. A. Prodi, L. Quintieri, V. Re, A. Rocchi, F. Ronga, F. Salemi, G. Soranzo, R. Sturani, L. Taffarello, R. Terenzi, G. Torrioli, R. Vaccarone, G. Vandoni, G. Vedovato, A. Vinante, M. Visco, S. Vitale, J. Weaver,  J. P. Zendri, P. Zhang, Results of the IGEC-2 search for gravitational wave bursts during 2005, (Physical Review D 76, 102001 (2007)
    \item P. Astone,  M. Bassan, E. Coccia, S. D’Antonio, V. Fafone, G. Giordano, A. Marini, Y. Minenkov, I. Modena, A. Moleti, G.V. Pallottino, G. Pizzella, A. Rocchi, F. Ronga, R. Terenzi, M. Visco, Analysis of 3 years of data from the gravitational wave detectors EXPLORER and NAUTILUS, (Physical Review D 87, 082002, 2013)
  
\end{enumerate}

This last paper is the most significant, highlighting the results of the overall activities.

A premonition of the activities of LIGO-Virgo was presented at our IRAP PhD and Erasmus Mundus School/Workshop, held in Les Houches from April 3 to 8, 2011, by Christian Ott. To the surprise of all the participants, he presented the new program of “blind injection” technique on an event of September 16, 2010. We found on Phys. Rev. D 95, 062002 (2017) Validating gravitational-wave detections: The Advanced LIGO hardware injection system” by a very large number of authors, implying a vast dedicated effort in this “very interesting, if true” approach. It is appropriate to recall the use of the term “interesting” by Niels Bohr, in reference to a wrong paper, this attitude was further expanded in the expression “very interesting, if true” by Eugene Wigner in Princeton (R. Ruffini 2025).

We were very surprised of the first detected observation of a GW signal on September 14, 2015 which for many reasons, in our opinion, was an injected event (see Ruffini R., 17th Italo-Korean meeting AIP Conf. Proc. 2874, 020014 (2024).

The same conclusion we reached on GW 170817 and GRB 170817A. The reason of the injected nature of GW 170817 was naturally reached by comparison with the nature of GRB 190510 (1) R. Ruffini et al., GRB 090510: a genuine short GRB from a binary neutron star coalescing into a Kerr–Newman Black Hole, ApJ Vol. 831, N. 2 (2016); see also 2) J.A. Rueda, R. Ruffini, Y. Wang, Short GRB 090510: A magnetized neutron star binary merger leading to a black hole, JHEAP, Vol. 50, 100464, 2026).

Similar conclusions appear to apply now to GRB 250114A, following the considerations in A. G. Abac et al., Testing Hawking’s Area Law and the Kerr Nature of Black Holes, Physical Review Letters 135, 111403 (2025), taking into account the 3 ICRANet papers (1) R. Ruffini et al., Single versus the Repetitive Penrose Process in a Kerr Black Hole, Phys. Rev. Letters 134, 081403 (2025); 2) Ruffini et al., Role of the irreducible mass in repetitive Penrose energy extraction processes in a Kerr black hole, Phys Rev Research 7, 013203 (2025) and 3) Ruffini R., Zhang S., in preparation).

Apart these 3 events with large signals to noise ratio, the remaining events follow the same problematic of the Weber events, which occur very often, the identification of GW template is vitiated by an inadequate analysis of thermal noise of the detectors. In addition, no astrophysical counterpart was found in the above 3 cases, nor in all the remaining cases of LIGO-Virgo-KAGRA, duplicating an inconsistency already present in Weber’s data.

Ruffini recalls that when discussing with Penrose the most recent understanding of the role of the irreducible mass, Penrose remarked that if everything were understood, then one should be able to explain the nature of quasars, a question he, Penrose, had unsuccessfully pursued for decades. Ruffini replied: yes, this can be addressed by considering a degenerate dark matter core accreting in an overdense region of our Universe (see 1) Ruffini R, Wang Y., Islands of Electromagnetic Tranquility in Our Galactic core and Little Red Dots that Shelter Molecules and Prebiotic Chemistrysubmitted for publication in ApJL, https://arxiv.org/abs/2509.17484v1 and 2) Wang, Y., \& Ruffini, R., Growth of High-Redshift Quasars from Fermion Dark Matter Seeds, submitted to JHEAP, 2025.).
\newpage
\bibliographystyle{ws-procs961x669}

\end{document}